\documentclass[aps, prb, reprint, groupedaddress, showkeys]{revtex4-2}

\usepackage{amsmath}
\usepackage{amssymb}
\usepackage{graphicx}
\usepackage[english]{babel}
\usepackage{orcidlink}

\DeclareMathOperator{\Real}{Re}
\DeclareMathOperator{\Imag}{Im}
\DeclareMathOperator{\sgn}{sgn}
\DeclareMathOperator{\Tr}{Tr}
\DeclareMathOperator{\Ln}{Ln}

\newcommand{\Hc}{\text{H.c.}}
\newcommand{\TO}{\mathcal{T}}
\newcommand{\ket}[1]{\left| #1 \right \rangle}
\newcommand{\nket}[1]{| #1 \rangle} 
\newcommand{\bra}[1]{\left \langle #1 \right |}
\newcommand{\kb}[2]{\left| #1 \middle \rangle  \middle \langle #2 \right|}

\newcommand{\bOk}[3]{\left\langle #1 \middle | #2 \middle | #3 \right \rangle}

\begin{document}

\title{Leading-logarithmic approximation by one-loop renormalization group
within Matsubara formalism}

\author{Jan Diekmann\,\orcidlink{0000-0002-1206-2407}}
\author{Severin G. Jakobs\,\orcidlink{0000-0002-0125-6353}}

\affiliation{Institut f\"ur Theorie der Statistischen Physik, RWTH Aachen
University, 52056 Aachen, Germany\\
and JARA---Fundamentals of Future Information Technology, 52056 Aachen, Germany}

\date{October 31, 2023}


\begin{abstract}
   We demonstrate how to devise a Matsubara-formalism-based one-loop
   approximation to the flow of the functional renormalization group (FRG) that
   reproduces identically the leading-logarithmic parquet approximation.  This
   construction of a controlled fermionic FRG approximation in a regime not
   accessible by perturbation theory generalizes a previous study from the
   real-time zero-temperature formalism to the Matsubara formalism and thus to
   the \emph{de facto} standard framework used for condensed-matter FRG studies.
   Our investigation is based on a simple model for the absorption of x rays in
   metals.  It is a core part of our construction to exploit that in a suitable
   leading-logarithmic approximation the values of the particle-hole
   susceptibility on the real- and on the imaginary-frequency axes are
   identical.
\end{abstract}

\maketitle


\section{Introduction}

Consider the function $x \mapsto x^{f(g)}$ in which the exponent $f(g) = a_1 g +
a_2 g^2 + \dots$ has a well-behaved power-series expansion without constant term
$a_0 g^0$.  Then
\begin{align}
   x^{f(g)} 
   &=
   \exp\left[(a_1 g + a_2 g^2 + \dots)\ln x\right]
   \\
   &= 
   x^{a_1 g} \exp\left[(a_2 g^2 + \dots)\ln x\right]
   \\
   &\approx
    x^{a_1 g}.
\end{align}
This approximation is good if $g\ll 1$ is so small that $g^2 \ln x \ll 1$. The
approximation allows even for large $\ln x$, and for $g \ln x$ that is not
small. In a clearly defined regime of $x$ the approximation is thus \emph{a
priori} known to be good, and it is known to become better with decreasing $g$.

Such a \emph{controlled} approximation is just the result of the
leading-logarithmic parquet approximation to certain many-body models in
condensed-matter theory whose perturbation theory suffers from logarithmic
divergencies.  In the above example, the failure of perturbation theory becomes
visible when one expands $x^{f(g)}$ up to first (or higher) order in $g$ (which
plays the role of the coupling constant),
\begin{equation}
   x^{f(g)} 
   \approx 
   1 + a_1 g \ln x.
\end{equation}
This approximation is indeed not appropriate when $g \ln x$ is not small.  Among
the most renowned applications of the leading-logarithmic parquet approximation
in condensed-matter physics are those to the Kondo model~\cite{Abrikosov1965},
to the one-dimensional interacting Fermi gas~\cite{Bychkov1966}, and to x-ray
absorption in metals~\cite{Roulet1969} from which the above example is taken.
These models of interacting zero- and one-dimensional Fermi systems share the
pattern in which logarithmic divergencies arise in diagrammatic perturbation
theory: the dominant, ``leading'' terms are contained in the so-called parquet
diagrams.

In Ref.~\cite{Diekmann2021} we demonstrated a close relation between the
leading-logarithmic parquet approximation and another, at first sight separate,
approximation: the one-loop flow of the functional renormalization group (FRG).
We substantiated for a model of x-ray absorption that a suitably crafted
one-loop FRG approximation is identical on a detailed technical level to the
leading-logarithmic parquet approximation.  While this complete identity of the
two methods is a unique observation, it has been known for a long time that
renormalization group (RG) approaches are able to reproduce the results of
leading-logarithmic parquet treatments:  This can be achieved already with basic
scaling techniques for the Kondo problem~\cite{Anderson1970}, for x-ray
absorption~\cite{Solyom1974} and for one-dimensional
conductors~\cite{Solyom1979} (see also~\cite{Bourbonnais1991}).

The FRG mentioned above is a modern formulation of the RG idea, based on
generating
functionals~\cite{Metzner2012,Kopietz2010,Platt2013,Salmhofer2019,Dupuis2021}.
In this formalism, the flow equation of a suitable generating functional entails
a hierarchy of flow equations for vertex functions.  The RG flow of the vertex
functions can also be understood from a distinctly diagrammatic point of
view~\cite{Jakobs2007}. This allows to compare FRG approximations in detail with
diagrammatic ones like the parquet approximation, as done, e.g., in
Refs.~\cite{Kugler2018, Kugler2018a, Kugler2018b, Diekmann2021}.

The FRG is nowadays applied to a broad variety of models in condensed-matter
theory, including interacting zero- and one-dimensional fermionic models for
quantum dots and wires~\cite{Metzner2012,Camacho2022}.  For some of these
applications, the applied approximations are known to reproduce correctly the
leading scaling behavior, which is not accessible by finite order perturbation
theory. Notable examples are the scaling of the back-scattering self-energy
component for Luttinger liquids with impurities~\cite{Meden2002} and the scaling
of the hopping between localized level and leads for the interacting resonant
level model at large bandwidth~\cite{Karrasch2010, Kennes2013}.  In these
particular cases, the leading approximation to the scaling behavior follows
already in the lowest level truncation scheme and allows thus for a short and
lucid derivation.  Many applications of the fermionic FRG to quantum dot
structures, however, do not focus on the idea of constructing a controlled
approximation, especially if they use a higher level truncation scheme. Instead,
the arguments applied to justify approximate truncations of the FRG flow
equations are often either perturbative or they are given \emph{a posteriori} by
comparing the results with those of other methods.  A characteristic example for
this situation is given by numerous FRG treatments~\cite{Hedden2004, Gezzi2007,
Karrasch2008, Bartosch2009,Isidori2010,Jakobs2010, Kinza2013, Streib2013,
Rentrop2016, Ge2023} of the single-impurity Anderson model~\cite{Anderson1961}.
It is known that the Anderson model at small $s$-$d$ mixing can be mapped onto
the antiferromagnetic spin-$\frac{1}{2}$ Kondo model by a unitary
transformation~\cite{Schrieffer1966}; and it is known that the summation of the
leading logarithms of the Kondo model provides meaningful results for the
scattering amplitude at temperatures moderately above the Kondo
temperature~\cite{Abrikosov1965}.  However, typical FRG treatments of the
Anderson model are not concerned with identifying these leading logarithms in a
suitable perturbation theory of that model in order to systematically include
them into the approximation.  Instead, most approaches start from integrating
out the lead propagation and treating the on-site interaction as perturbation;
this leads to a perturbation theory without logarithmic divergencies.  The FRG
approaches then typically achieve a somewhat extended range of applicability
compared to second-order perturbation theory, but do not provide a reliable
access to the Kondo regime~\cite{Hedden2004, Gezzi2007, Karrasch2008,
Bartosch2009,Isidori2010,Jakobs2010, Rentrop2016, Ge2023}.  One FRG scheme
exploiting Ward identities was able to reproduce Kondo scaling but used four fit
parameters for an interaction-dependent regulator~\cite{Streib2013}. And while
Ref.~\cite{Kinza2013} is exceptional in being based on a differently organized
perturbation theory, it does not provide a systematic consideration of leading
contributions.  In summary, none of the cited FRG approaches to the Anderson
impurity model are controlled \emph{a priori} in the Kondo regime.

We believe that it is worthwhile to reinforce the quest for controlled FRG
approximations applicable to models for quantum dots and wires, in particular
for situations which require truncations beyond the one on the lowest level.
Such approximations could combine the efficiency and analytic accessibility of
the FRG with an \emph{a priori} known reliability.  As a first step in this
program we constructed in Ref.~\cite{Diekmann2021} the aforementioned one-loop
FRG approximation that reproduces identically the leading-logarithmic parquet
approximation for a model of x-ray absorption in metals.  In that reference, we
created and used a formulation of the FRG within the real-time zero-temperature
ground-state formalism to be as close as possible to the parquet treatment of
this model from Ref.~\cite{Roulet1969}.  However, most applications of the FRG
to condensed-matter problems are implemented within the imaginary-time
finite-temperature Matsubara
formalism~~\cite{Metzner2012,Kopietz2010,Platt2013,Salmhofer2019,Dupuis2021}. It
is hence naturally our next goal to transfer our analysis of
Ref.~\cite{Diekmann2021} to the Matsubara formalism.  This transfer is realized
in this paper:  We show how to devise, within Matsubara formalism, a
one-loop FRG approximation for the x-ray absorption problem that reproduces
identically the leading-logarithmic parquet approximation.

The paper is organized as follows.  In Sec.~\ref{sec:X-ray-absorption} we set up
the formal framework to describe x-ray absorption in metals.  We start by
briefly introducing the underlying model in Sec.~\ref{subsec:Model-system}. Then
we discuss in Sec.~\ref{subsec:Ground-state} how the ground state of that system
depends on the energy level of the localized deep state.  On this basis we can
later relate the description of the model in the Matsubara formalism to the one
in the real-time zero-temperature ground-state formalism.  The relation between
the formalisms then helps us to benefit from results of Refs.~\cite{Roulet1969,
Diekmann2021}.  In Sec.~\ref{subsec:Linear-response-rate} we delineate how the
linear response rate of x-ray absorption can be computed from the imaginary-time
exciton propagator. This propagator in turn is accessible by fermionic
diagrammatic perturbation theory in Matsubara formalism.  In
Sec.~\ref{sec:Specifications-of-perturbation-theory} we expound how we employ
the Matsubara formalism at $T=0$ with continuous Matsubara frequencies and with
a specific choice of the deep-state energy.  Within this formal framework we
then apply a leading-logarithmic approximation to the local conduction-state
propagator in Sec.~\ref{sec:Leading-log-approx}.  We show that this
approximation makes the imaginary-frequency diagrammatic expressions coincide
with the real-frequency ones.  From that observation onwards all further steps
can be copied from our previous study in Ref.~\cite{Diekmann2021}.  In
Sec.~\ref{sec:Comparison-to-Kugler} we relate our construction of a
leading-logarithmic one-loop Matsubara FRG approximation to the observation from
Ref.~\cite{Kugler2018a} that multiloop FRG improves the results of one-loop FRG
for the model under consideration.  Finally, we conclude in
Sec.~\ref{sec:Conclusion}.  Detailed explanations of several technical issues
can be found in the Appendixes.


\section{X-ray absorption and exciton propagator}
\label{sec:X-ray-absorption}

In this section we establish the formal basis underlying all later
considerations.  First we introduce (Sec.~\ref{subsec:Model-system}) a simple
model system for x-ray absorption known from Ref.~\cite{Roulet1969}.  Later we
identify the propagator of a local exciton as a many-body quantity that allows
to determine the rate of x-ray absorption
(Sec.~\ref{subsec:Linear-response-rate}).  In order to relate the Matsubara
exciton propagator to the one from the real-time zero-temperature formalism, a
clear understanding of the ground state of the model system is required.
Therefore, the ground state is studied in Sec.~\ref{subsec:Ground-state}.


\subsection{Model system}
\label{subsec:Model-system}

The model studied in this paper is identical to the one in our previous
paper~\cite{Diekmann2021} and essentially taken from Ref.~\cite{Roulet1969}.  It
is a basic model for the theoretical analysis of x-ray absorption in metals.
The Hamiltonian of spinless electrons is given by
\begin{align}
    H 
    &= 
    H_{0} - \frac{U}{V} \sum_{kk'} a_{k'}^{\dagger} a_{k}
      a_{\text{d}} a_{\text{d}}^{\dagger}, 
    \label{eq:Hamiltonian}
    \\
    H_{0} 
    & =
    \epsilon_{\text{d}}a_{\text{d}}^{\dagger}a_{\text{d}} 
    + \sum_{k} \epsilon_{k} a_{k}^{\dagger} a_{k}. 
    \label{eq:H0}
\end{align}
Here, $a_{\text{d}}^{\dagger}$ creates an electron in a localized ``deep'' state
with energy $\epsilon_{\text{d}} < 0$ and $a_{k}^{\dagger}$ creates an electron
with momentum $k$ in the conduction band which extends from $-\xi_{0}$ to
$\xi_{0} > 0$ with a constant density of states $\rho$.  For simplicity, we
assume the total number of momentum eigenstates in the conduction band to be
even and denote it by $2N = 2\xi_{0}\rho$.  A hole on the deep state leads to a
local attractive potential for the electrons in the conduction band.  The
amplitude $-U < 0$ of this interaction is independent of the conduction-electron
momenta; $V$ denotes the volume of the conductor.  We consider the system
primarily at zero temperature and with a half-filled conduction band.

Our quantity of interest is the linear response rate $R(\nu)$ of absorption of
x-rays from a perturbing x-ray field with frequency $\nu$.  This field is
assumed to interact with the system via an additional addend to the Hamiltonian,
namely,
\begin{equation}
    H_{X}(t) = e^{-i\nu t} W A^{\dagger} + \Hc
\end{equation}
Here,
\begin{equation}
    A^{\dagger} = \frac{1}{\sqrt{V}} \sum_{k} a_{k}^{\dagger} a_{\text{d}}
\end{equation}
creates a local particle-hole excitation.  As the electromagnetic field is not
quantized, the rate of x-ray absorptions is derived from the rate of
electronic transitions (compare Sec.~\ref{subsec:Linear-response-rate}). 

The absorption rate is known to vanish for frequencies below some threshold
$\nu_{c}$, above which the leading behavior is a power-law divergence $R(\nu)
\propto [\xi_{0}/(\nu-\nu_{c})]^{2g}$ with $g=\rho U/V$ denoting the
dimensionless coupling constant~\cite{Roulet1969, Nozieres1969, Nozieres1969a}.
An expansion of $R(\nu)$ in powers of $g$ leads to a series in powers of
$g\ln[\xi_{0}/(\nu-\nu_{c})]$.  (We use units with $\hbar=1$ and also $k_\text
B=1$.)  Correspondingly, a treatment of this model with many-body perturbation
theory leads to logarithmic divergencies; these appear in bubbles of
particle-particle and particle-hole propagation.  The way in which the
divergencies occur is prototypical for a class of low-dimensional fermionic
systems which includes the Kondo model~\cite{Abrikosov1965} and the
one-dimensional interacting Fermi gas~\cite{Bychkov1966, Solyom1979}.


\subsection{Ground state of the system}
\label{subsec:Ground-state}

In the subsequent sections it is necessary to understand how the ground state of
the system filled with $N+1$ particles depends on the value of
$\epsilon_{\text{d}}$.  This is clarified in the present section in which we
explain that there exist two many-body states $\ket{\Psi_{0}}$ and
$\nket{\bar{\Psi}_{0}}$ and a threshold value $\epsilon_{\text{d}0}$ such that
the ground state changes from $\ket{\Psi_{0}}$ for
$\epsilon_{\text{d}}<\epsilon_{\text{d}0}$ to $\nket{\bar{\Psi}_{0}}$ for
$\epsilon_{\text{d}}>\epsilon_{\text{d}0}$. (The particular case
$\epsilon_{\text{d}}=\epsilon_{\text{d}0}$ with a degeneracy of the ground state
will not be important for us.)

Since the Hamiltonian commutes with the deep-state occupancy $n_{\text{d}}$
there exists a Hilbert space basis that consists of common eigenstates of both
operators: the eigenstates of $H$ can be chosen to have either an occupied or an
empty deep level.  Let $\ket{\Psi_{0}}$ denote the energetically lowest state in
the subspace with occupied deep level and let $\nket{\bar{\Psi}_{0}}$ denote the
energetically lowest state in the subspace with empty deep level.  Depending on
$\epsilon_{\text{d}}$, either $\ket{\Psi_{0}}$ or $\nket{\bar{\Psi}_{0}}$ has
the lower energy and is thus the ground state.

We first consider $\ket{\Psi_{0}}$.  Since $H$ equals $H_{0}$ on the subspace
with $n_{\text{d}}=1$, $\ket{\Psi_{0}}$ is the Slater determinant type many-body
state in which the deep level and the $N$ momentum states in the lower half of
the conduction band are occupied, while the states in the upper half of the
conduction band are empty. The energy of this state is
\begin{equation}
   E_{0} = \epsilon_{\text{d}} + E_{\text{cb}}
\end{equation}
where $E_{\text{cb}}$ is the energy of the half filled conduction band,
\begin{equation}
   E_{\text{cb}}
   =
   \rho \int_{-\xi_{0}}^{0} d\epsilon \, \epsilon = -\frac{\rho\xi_{0}^{2}}{2}.
\end{equation}

Next we discuss $\nket{\bar{\Psi}_{0}}$.  On the subspace with $n_{\text{d}}=0$,
the Hamiltonian $H$ is equal to the perturbed single-particle Hamiltonian 
\begin{equation}
   \bar{H}_{0}
   =
   \sum_{k,k'} \left( \delta_{kk'}\epsilon_{k} - \frac{U}{V} \right)
    a_{k'}^{\dagger} a_{k}.
\end{equation}
The single-particle eigenstates of $\bar{H}_{0}$ are not the momentum states
$k_{n}$ but perturbed scattering states $\bar{k}_{n}$ to energies
$\bar{\epsilon}_{n}$, which we sort as $\bar{\epsilon}_{1} < \bar{\epsilon}_{2}
< \dots < \bar{\epsilon}_{2N}$.  To be more precise, the particular state with
the lowest perturbed single-particle energy $\bar{\epsilon}_{1} =:
\bar{\epsilon}_{\text{b}}$ is a bound state resulting from the localized
attractive potential generated by the deep hole.  Its energy is approximately
\begin{equation}
   \bar{\epsilon}_{\text{b}} \approx -\xi_{0} \left( 1 + 2e^{-1/g} \right)
\end{equation}
for $1 \gg g \gg 1/\ln(2\xi_{0}\rho)$ (see
Appendix~\ref{sec:App-Details-on-ground-state} for more details).  In the
$(N+1)$-particle state $\nket{\bar{\Psi}_{0}}$, the bound state and the
scattering states $\bar{k}_{2},\dots,\bar{k}_{N+1}$ are occupied while the
states $\bar{k}_{N+2},\dots\bar{k}_{2N}$ are empty. The energy of
$\nket{\bar{\Psi}_{0}}$ is thus
\begin{equation}
   \bar{E}_{0} = \bar{\epsilon}_{\text{b}} + \bar{E}_{\text{cb}},
\end{equation}
where
\begin{equation}
   \bar{E}_{\text{cb}} = \sum_{n=2}^{N+1} \bar{\epsilon}_{n}
\end{equation}
is the energy of the half filled band of scattering states. 

From the comparison of $E_0$ and $\bar E_0$ follows that there is a threshold
value $\epsilon_{\text{d}0}$ given by the condition
\begin{equation}
   \epsilon_{\text{d}0} + E_{\text{cb}} 
   = 
   \bar{\epsilon}_{\text{b}} + \bar{E}_{\text{cb}},
\end{equation}
such that $\ket{\Psi_{0}}$ is the ground state for $\epsilon_{\text{d}} <
\epsilon_{\text{d0}}$ while $\nket{\bar{\Psi}_{0}}$ is the ground state for
$\epsilon_{\text{d}} > \epsilon_{\text{d0}}$.  It can be shown that
$E_{\text{cb}} < \bar{E}_{\text{cb}} < E_{\text{cb}}+\xi_{0}$ (see
Appendix~\ref{sec:App-Details-on-ground-state}).  As a consequence,
$\epsilon_{\text{d}0}$ is strictly negative.

The deep state of the above model for x-ray absorption in metals describes an
electron which is localized close to the core and which fully screens the core
from the conduction electrons.  The energy gain of binding such a deep electron
should be greater than the gain that results from exposing the delocalized
conduction electrons to the core charge.  Therefore, the situation of physical
interest is given by $\epsilon_{\text{d}} < \epsilon_{\text{d}0}$, with
$\ket{\Psi_{0}}$ being the ground state.  In
Sec.~\ref{sec:Specifications-of-perturbation-theory} we will find that it is of
technical advantage to study the system at $\tilde{\epsilon}_{\text{d}} = 0^{-}$
(with $\tilde{\epsilon}_{\text{d}} = \epsilon_{\text{d}} + g\xi_{0}$ denoting
the Hartree renormalized deep-state energy) and that the corresponding results
can be directly related to those for physically relevant values of
$\epsilon_{\text{d}}$.


\subsection{Linear response rate and exciton propagators}
\label{subsec:Linear-response-rate}

In this section, we summarize how the linear response rate of x-ray absorption
can be computed from the retarded exciton propagator, and how the retarded, time-ordered and imaginary-time Matsubara exciton propagators are related. This
provides the foundation for the Matsubara-formalism-based approach to the
absorption rate in subsequent sections.

We suppose that the state of the system at some time $t_{0}$ is described by a
density operator $\varrho$ (not to be confused with the density of states $\rho$
in the conduction band) that commutes with $H$ and $n_\text{d}$.  Later we will
focus on the cases $\varrho = \kb{\Psi_{0}}{\Psi_{0}}$ and $\varrho = e^{-\beta
H}/Z$.  From $t_{0}$ on, the system evolves under the Hamiltonian
$H_{\text{tot}}(t) = H + H_{X}(t)$. During that time, $H_{X}(t)$ induces
transitions between the deep state and the conduction band such that the mean
deep-state occupancy changes. The linear response rate of this change approaches
a long time limit,
\begin{equation}
   \frac{d}{dt} \langle n_{\text{d}} \rangle(t)
   \xrightarrow{t_{0} \rightarrow -\infty}
   2 |W|^{2} \Imag \chi^{\text{ret}}(\nu).
\end{equation}
Here, $\nu$ denotes the x-ray frequency and
\begin{align}
   \chi^{\text{ret}}(\nu) 
   & =
   \int_{-\infty}^{\infty} dt\, e^{i \nu t} \chi^{\text{ret}}(t),
   \\
   \chi^{\text{ret}}(t) 
   & =
   -i \Theta(t) \left \langle \left[A(t)_{H}, A^{\dagger}\right] \right \rangle 
\end{align}
denotes the retarded local exciton propagator, with $A(t)_{H} =
e^{iHt}Ae^{-iHt}$.  Details on the derivation of this standard result are given
in Appendix~\ref{sec:App-linear-response-rate}.  

The above value $2|W|^{2} \Imag \chi^{\text{ret}}(\nu)$ is a good approximation
for $d\langle n_{\text{d}} \rangle / dt$ in a certain time regime only: The time
span $(t-t_{0})$ must be sufficiently long for the linear response rate to
approach its long-time limit, and it must be sufficiently short for the linear
response approximation to be applicable. The second condition restricts the
interval of allowed times by an upper boundary which can be pushed to higher
values by decreasing the amplitude $W$ of the perturbation. 

Concerning the relation between the retarded and the Matsubara exciton
propagator, it will be important later in this section that
$\chi^{\text{ret}}(z)$ is analytic in the open upper half of the complex $z$
plane.  This region of analyticity of $\chi^{\text{ret}}(z)$ can be established
in the usual way by use of the Lehmann representation (see
Appendix~\ref{sec:App-analytic-behaviour}).  We use $z$ instead of $\nu$ to
denote complex frequency arguments.  The value of $\chi^{\text{ret}}(z)$ at real
$z=\nu$ is given by the limit of vanishing imaginary part from above,
$\chi^{\text{ret}}(\nu) \equiv\chi^{\text{ret}}(\nu + i \eta)$.

Let use now suppose that the system is initially in the state $\varrho =
\kb{\Psi_{0}}{\Psi_{0}}$.  Then one can use $\bra{\Psi_{0}} A^{\dagger} = 0$ to
show that the retarded exciton propagator $\chi_{\Psi_{0}}^{\text{ret}}(t)$
resulting from that initial state is (for $t \neq 0$) identical to the
time-ordered one $\chi_{\Psi_{0}}(t)$:
\begin{align}
   \chi_{\Psi_{0}}^{\text{ret}}(t)
   & =
   -i \Theta(t) \bOk{\Psi_{0}}{[A(t)_{H}, A^{\dagger}]}{\Psi_{0}}
   \\
   & =
   -i \Theta(t) \bOk{\Psi_{0}}{A(t)_{H} A^{\dagger}}{\Psi_{0}}
   \\
   & =
   -i \bOk{\Psi_{0}}{\TO A(t)_{H} A^{\dagger}}{\Psi_{0}}
   \\
   & =:
   \chi_{\Psi_{0}}(t).
\end{align}
Here, the time-ordering operator $\TO$ rearranges $A(t)_{H}$ and $A^{\dagger} =
A^{\dagger}(0)_H$ such that their time arguments decrease from left to right.
The time-ordered exciton propagator in frequency representation,
\begin{equation}
   \chi_{\Psi_{0}}(\nu)
   =
   \int_{-\infty}^{\infty} dt\, e^{i\nu t} \chi_{\Psi_{0}}(t),
\end{equation}
is by definition identical to the particle-hole susceptibility studied in our
previous paper~\cite{Diekmann2021}; compare also the similar susceptibilities
studied in~\cite{Roulet1969, Nozieres1969, Kugler2018, Lange2015, Kugler2018a}.
For $\epsilon_{\text{d}} < \epsilon_{\text{d}0}$, the state $\ket{\Psi_{0}}$ is
the ground state. Then
\begin{equation}
   -\frac{d}{dt} \langle n_{\text{d}} \rangle (t)
   \xrightarrow{t_{0} \rightarrow -\infty}
   -2 |W|^{2} \Imag \chi_{\Psi_{0}}(\nu)
   =
   R(\nu)
\end{equation}
is the rate of x-ray induced excitations from the ground state, i.e., the rate
of x-ray absorption (compare Ref.~\cite{Roulet1969}).

Next we consider the case that the system is prepared initially in the thermal
density operator $\varrho = e^{-\beta H}/Z$ with temperature $T = 1/\beta \neq
0$, chemical potential $\mu = 0$ and with $Z = \Tr e^{-\beta H}$.  For $\tau \in
(-\beta, \beta)$ define the imaginary-time Matsubara exciton propagator
\begin{equation}
   \chi_{\beta}^{\text{Mat}}(\tau)
   =
   -\Tr\frac{e^{-\beta H}}{Z} \TO A(\tau)_{H} A^{\dagger},
\end{equation}
with $A(\tau)_{H} = e^{H\tau} A e^{-H\tau}$.  Here, $\TO$ sorts operators with
larger imaginary time $\tau$ to the left. For bosonic Matsubara frequencies
$X_{n} = 2 n \pi / \beta$ define furthermore
\begin{equation}
   \chi_{\beta}^{\text{Mat}}(iX_{n})
   =
   \int_{0}^{\beta} d\tau \, e^{iX_{n}\tau} \chi_{\beta}^{\text{Mat}}(\tau).
   \label{eq:chi^Mat(iXn)}
\end{equation}
As expected, the Lehmann representation of $\chi_{\beta}^{\text{Mat}}(iX_{n})$
is found to coincide with that of $\chi_{\beta}^{\text{ret}}(z\!=\!iX_{n})$ for
positive $X_{n}$ (see Appendix~\ref{sec:App-analytic-behaviour}).

\begin{figure}
   \begin{center}
      \includegraphics[width=7cm]{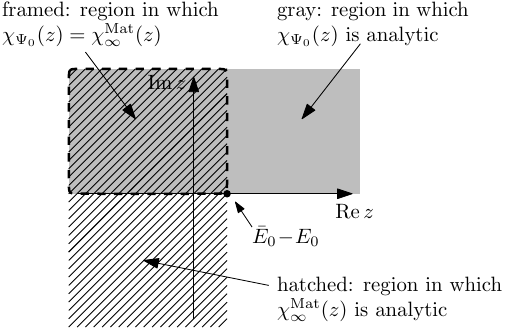}
   \end{center}
   \caption{Domains of analyticity of $\chi_{\infty}^{\text{Mat}}(z)$ from
      Eq.~(\ref{eq:chi^Mat:tau-to-nu-1-0}) and
      $\chi_{\Psi_{0}}(z)=\chi^{\text{ret}}_{\Psi_0}(z)$ from
      Eq.~(\ref{eq:chi-ret-Lehmann-0}) (in the limit $\beta \rightarrow \infty$)
      in the case $\epsilon_{\text{d}} < \epsilon_{\text{d}0}$ in which
      $\ket{\Psi_{0}}$ is the ground state.}
   \label{fig:Domains-of-analyticity1}
\end{figure}

In the limit $\beta \rightarrow \infty$ of vanishing temperature,
$\chi_{\beta}^{\text{Mat}}(iX_{n})$ approaches a function
$\chi_{\infty}^{\text{Mat}}(iX)$ which is defined on the whole
imaginary-frequency axis.  The analytic properties of this function are derived
in Appendix~\ref{sec:App-analytic-behaviour}.  They can be summarized as
follows: In the case $\epsilon_{\text{d}} < \epsilon_{\text{d}0}$ in which
$\ket{\Psi_{0}}$ is the ground state, the definition of
$\chi_{\infty}^{\text{Mat}}(z\!=\!iX)$ can be extended to the region given by
$\Real z < \bar{E}_{0} - E_{0}$, which includes the left half plane of $z$.  On
the intersection of this domain with the upper half plane,
$\chi_{\infty}^{\text{Mat}}(z)$ coincides with $\chi_{\Psi_{0}}(z) =
\chi_{\Psi_{0}}^{\text{ret}}(z)$.  This makes an analytic continuation from
$\chi_{\infty}^{\text{Mat}}(iX)$ to $\chi_{\Psi_{0}}(\nu)$ possible.  This
relation between $\chi_{\infty}^{\text{Mat}}(z)$ and $\chi_{\Psi_{0}}(z)$ is
sketched in Fig.~\ref{fig:Domains-of-analyticity1}.  We remark in passing that
in the case $\epsilon_{\text{d}} > \epsilon_{\text{d}0}$ in which
$\nket{\bar{\Psi}_{0}}$ is the ground state, $\chi_{\infty}^{\text{Mat}}(iX)$
can be analytically continued to the time ordered exciton propagator
$\chi_{\bar{\Psi}_{0}}(\nu)$ of the system prepared initially in the state
$\ket{\bar{\Psi}_{0}}$.

Later we will find an even simpler relation between $\chi_{\infty}^{\text{Mat}}$
and $\chi_{\Psi_{0}}$: For an appropriate choice of $\epsilon_\text{d}$ and in
leading-logarithmic approximation holds indeed $\chi_{\infty}^{\text{Mat}}(iX) =
\chi_{\Psi_{0}}(X)$ (compare Sec.~\ref{sec:identity-leading-log-real-imag}).


\section{Specifications of perturbation theory in Matsubara formalism}  
\label{sec:Specifications-of-perturbation-theory}

In this section we specify details of our use of the Matsubara formalism which
are crucial for the construction of the leading-logarithmic one-loop FRG
approximation in Sec.~\ref{sec:Leading-log-approx}.  While the appearance of
fermionic two-particle functions in
Sec.~\ref{subsec:Appearance-of-fermionic-MGF} and the use of Hartree dressed
propagators in Sec.~\ref{subsec:Hartree-dressed-propagators} are completely
analogous to our prior treatment in the real-time zero-temperature formalism in
Ref.~\cite{Diekmann2021}, a major difference is that temperature and $\tilde
\epsilon_\text{d}$ now regularize the logarithmic divergencies
(Sec.~\ref{subsec:Regularizing-effects}).  We restrict our considerations to
vanishing temperature.  In Sec.~\ref{subsec:Zero-temperature-limit} we explain
that the particular parquet diagrams involved then allow us to pass over to
continuous Matsubara frequencies.  Technically we are then so close to the
real-time zero-temperature formalism (compare
Appendix~\ref{sec:App-identification-Matsubara-0T}) that we can exploit a Ward
identity proven in that formalism in Ref.~\cite{Diekmann2021} to set $\tilde
\epsilon_\text{d}$ to $0^-$; this is explained in
Sec.~\ref{subsec:Setting-eps_d=0}.


\subsection{Appearance of fermionic Matsubara Green functions}
\label{subsec:Appearance-of-fermionic-MGF}

When we described the relation between $\chi_{\Psi_{0}}$ and
$\chi_{\beta}^{\text{Mat}}$ in the preceding section, we interpreted these
susceptibilities as propagators of a single bosonic excitation.  Now we aim at a
perturbative and, later on, RG approach to the susceptibilities; for that
purpose we treat them as fermionic two-particle functions.  In fact, it is
\begin{align}
   & \chi_{\beta}^{\text{Mat}}(iX_{n}) \nonumber
   \\
   & =
   \frac{1}{V} \int_{0}^{\beta} d\tau \, \sum_{kk'} e^{iX_{n}\tau} 
   G_{\beta}^{\text{Mat}}(\text{d},0;k,\tau|\text{d},\tau;k',0)
   \label{eq:chi-from-GF}
   \\
   & =
   \frac{1}{V\beta^{3}} 
   \sum_{\substack{n_{1} n'_{1} n'_{2} \\ k k'}}
   G_{\beta}^{\text{Mat}}(\text{d},\omega_{n_{1}};
   k,\omega_{n'_{1}}+X_{n}| \text{d},\omega_{n'_{1}};k',\omega_{n'_{2}}).
\end{align}
Here the fermionic Matsubara two-particle Green function and its representation
in terms of Matsubara frequencies are
\begin{align}
   & G_{\beta}^{\text{Mat}}(l_{1},\tau_{1};l_{2},\tau_{2}|
   l'_{1},\tau'_{1};l'_{2},\tau'_{2}) \nonumber
   \\
   &=
   (-1)^{2}\Tr\frac{e^{-\beta H}}{Z} \TO a_{l_{1}}(\tau_{1})_{H} 
   a_{l_{2}}(\tau_{2})_{H} a_{l'_{2}}^{\dagger}(\tau'_{2})_{H}
   a_{l'_{1}}^{\dagger}(\tau'_{1})_{H},
   \\
   & G_{\beta}^{\text{Mat}}(l_{1},\omega_{n_{1}};l_{2},\omega_{n_{2}}|
   l'_{1},\omega_{n'_{1}};l'_{2},\omega_{n'_{2}}) \nonumber
   \\
   & =\int_{0}^{\beta} d\tau_{1} \dots \int_{0}^{\beta} d\tau'_{2} \, 
   e^{i(\tau_{1}\omega_{n_1} + \dots - \tau'_{2}\omega_{n'_2})}
   \nonumber \\
   & \quad\;
   \times G_{\beta}^{\text{Mat}}(l_{1},\tau_{1};l_{2},\tau_{2}|
   l'_{1},\tau'_{1};l'_{2},\tau'_{2}).
   \label{eq:2-particle-Matsu-GF-freq}
\end{align}
In these equations, the $l_{i}$ are single-particle state indices, and time
ordering swaps by definition the sign of the expression if the resulting
sequence of fermionic ladder operators is an odd permutation of the initial one.
In Eq.~(\ref{eq:2-particle-Matsu-GF-freq}), the $\omega_{n}=(2n+1)\pi T$ with
integer $n$ denote fermionic Matsubara frequencies, and
$G_{\beta}^{\text{Mat}}(l_{1},\omega_{n_{1}};l_{2},\omega_{n_{2}}|
l'_{1},\omega_{n'_{1}};l'_{2},\omega_{n'_{2}})$ is proportionate to
$\beta\delta_{n_{1}+n_{2},n'_{1}+n'_{2}}$.

Standard imaginary-time perturbation theory yields diagrammatic approximations
for the fermionic Green functions. The values of the corresponding Hugenholtz
diagrams can be determined with the usual Matsubara diagram rules as
\begin{equation}
   \label{eq:diagram-rule}
   \frac{(-1)^{P}(-1)^{n_{\text{loop}}}}{2^{n_{\text{eq}}}S}
   \left[\prod\left(-\bar \nu\right)\right]
   \left[\prod g_{\beta}^{\text{Mat}}\right]
\end{equation}
(compare, e.g., Ref.~\cite{Negele1988}).  Here, the vertex $\bar \nu_{1'2'|12}$,
with indices $1=(l_1,\tau_1)$ etc., satisfies
\begin{equation}
   \bar \nu_{\text d \tau'_{1}, k' \tau'_{2} | \text{d} \tau_{1}, k \tau_{2}}
   =
   \frac{U}{V} \delta(\tau'_1-\tau'_2) \delta(\tau'_2-\tau_1) 
     \delta(\tau_1 - \tau_2). 
\end{equation}
Several further components of the vertex are determined through the antisymmetry
relation $\bar \nu_{1'2'|12} = \bar \nu_{2'1'|21} = -\bar \nu_{2'1'|12} = - \bar
\nu_{1'2'|21}$.  All other components of the vertex vanish.  In the diagram
rule~(\ref{eq:diagram-rule}), the products run over all vertices and all lines;
the latter represent free propagators $g_{\beta}^{\text{Mat}}$.  Implicitly, all
internal indices for states and times or frequencies are to be summed over.
$(-1)^{P}$ is defined as $(+1)$ if the first incoming line of the diagram is
connected to the first outgoing line and as $(-1)$ if it is connected to the
second outgoing line. $n_{\text{loop}}$ is the number of internal closed loops,
$n_{\text{eq}}$ is the number of pairs of equivalent lines, and $S$ is the
diagram symmetry factor.


\subsection{Hartree dressed propagators}
\label{subsec:Hartree-dressed-propagators}

The free propagator in Matsubara formalism at finite temperature is given in
time or frequency representation, respectively, by
\begin{align}
   g_{\beta}^{\text{Mat}}(\tau) 
   & =
   e^{-h\tau}\left[f(h)-\Theta(\tau-\eta)\right],
   \label{eq:free-Matsu-prop-tau}
   \\
   g_{\beta}^{\text{Mat}}(i\omega_{n}) 
   & =
   \frac{e^{i\omega_{n}\eta}}{i\omega_{n}-h}.
   \label{eq:free-Matsu-prop-omega_n}
\end{align}
These are equations for matrices. The matrix indices, not written here, refer to
single-particle states. In particular, $h$ is the matrix of single-particle
energies. Furthermore, $f(h)=\left(e^{\beta h}+1\right)^{-1}$ denotes the
(matrix) Fermi function at vanishing chemical potential;  $\Theta$ denotes the
Heaviside step function and $\eta$ is a positive infinitesimal.

The matrix $h$ does not only comprise the energies $\epsilon_{\text{d}}$ and
$\epsilon_{k}$ appearing in the Hamiltonian in Eq.(\ref{eq:H0}) but also a
single-particle perturbation that appears when the interaction term in
Eq.~(\ref{eq:Hamiltonian}) is brought into the standard form by permuting all
creation operators to the left.  At $T=0$, that perturbation is precisely
canceled by the Hartree self-energy.  This suggests to work with Hartree dressed
propagators, as we already did in our previous analysis of the model within the
zero-temperature formalism~\cite{Diekmann2021}.  Hartree dressed propagators
result when $h$ is replaced in Eqs.~(\ref{eq:free-Matsu-prop-tau}) and
(\ref{eq:free-Matsu-prop-omega_n}) by $h^{\text{H}}$ with components
\begin{align}
   h_{k'k}^{\text{H}} 
   & =
   \delta_{k'k}\epsilon_{k}-\left[1-f(\epsilon_{\text{d}})\right]\frac{U}{V},
   \\
   h_{\text{d}\text{d}}^{\text{H}} 
   & =
   \epsilon_{\text{d}}+g\xi_{0},
   \label{eq:deep-state-Hartree}
   \\
   h_{\text{d}k}^{\text{H}} & =0 = h_{k\text{d}}^{\text{H}}.
\end{align}
The contribution $-U/V$ to $h_{k'k}^{\text{H}}$ represents the aforementioned
single-particle perturbation, while $f(\epsilon_{\text{d}})U/V$ is the Hartree
self-energy. In the limit $T \rightarrow 0$, these two contributions cancel out
due to $f(\epsilon_{\text{d}}) \rightarrow 1$.  For moderate $T \ll \xi_{0}$ and
for typical $\epsilon_{\text{d}} < -\xi_{0}$ holds still $1 -
f(\epsilon_{\text{d}}) \lll 1$ and the eigenenergies $\tilde{\epsilon}_{n}$ of
$h_{k'k}^{\text{H}}$ can be reasonably approximated by first order perturbation
theory as $\tilde{\epsilon}_{n} = \epsilon_{n} - \left[1 -
f(\epsilon_{\text{d}})\right] U/V$. The first order shift 
\begin{equation}
   \left[1 - f(\epsilon_{\text{d}})\right] U/V 
   = \left[1 - f(\epsilon_{\text{d}})\right] g \xi_{0} / N \lll \xi_{0} / N
\end{equation}
is so small that these energies can still be described by a band with density
$\rho$ from $-\xi_{0}$ to $\xi_{0}$. 

The Hartree self-energy contribution to the deep-state energy in
Eq.~(\ref{eq:deep-state-Hartree}) follows from 
\begin{equation}
   \frac{U}{V} \rho \int_{-\xi_{0}}^{\xi_{0}} d\epsilon \, f(\epsilon)
   = g \xi_{0}.
\end{equation}
This value is independent of temperature due to $f(\epsilon) + f(-\epsilon)=1$
and due to the symmetry of the band around the chemical potential $\mu=0$.  We
define $\tilde{\epsilon}_{\text{d}} := \epsilon_{\text{d}}+g\xi_{0}$.  The
physically relevant case is $\tilde{\epsilon}_{\text{d}} < 0$.

In the following computation of two-particle quantities we will consider only
skeleton diagrams with Hartree dressed propagators, without any further
self-energy type contributions on the diagram lines.  This means that the
Hartree dressed propagators take the role of the full single-particle Green
functions. Therefore, we will denote the Hartree dressed propagator simply by
$G_{\beta}^{\text{Mat}}(\tau)$ or $G_{\beta}^{\text{Mat}}(i\omega_{n})$,
slightly abusing the usual notation for full propagators.


\subsection{Regularizing effect of temperature and
$\tilde{\epsilon}_{\text{d}}$}
\label{subsec:Regularizing-effects}

Within the real-time zero-temperature formalism, the archetype of a logarithmic
divergence arises in the zeroth order diagram (with Hartree dressed propagators)
to the particle-hole susceptibility.  That motivates us to study now the same
diagram within Matsubara formalism.  We find again a logarithmic divergence; it
is, however, regularized by temperature and by $\tilde{\epsilon}_{\text{d}}$.

\begin{figure}
   \begin{center}
      \includegraphics[width=2cm]{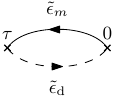}
   \end{center}
   \caption{Zeroth order diagram for $\chi^\text{Mat}_\beta(\tau)$.  The dashed
   line denotes a deep-state propagator and the solid line denotes a
   conduction-state propagator. An extra particle with energy $\tilde
   \epsilon_m$ in the conduction band and a hole at the deep-state energy
   $\tilde \epsilon_\text{d}$ propagate from $0$ to $\tau$.}
   \label{fig:Zeroth-order-diagram}
\end{figure}
The zeroth-order contribution to $\chi_{\beta}^{\text{Mat}}(iX_{n})$ at finite
temperature and $\epsilon_{\text{d}} < \epsilon_{\text{d}0}$ stems from the
diagram shown in Fig.~\ref{fig:Zeroth-order-diagram}.  It has the value
\begin{align} 
   & \chi_{0,\beta}^{\text{Mat}}(iX_{n}) \nonumber \\
   & =
   \int_{0}^{\beta} d\tau \, e^{iX_{n}\tau} \frac{1}{V} \sum_{m} 
   e^{(\tilde{\epsilon}_{\text{d}} - \tilde{\epsilon}_{m})\tau} 
   f(\tilde{\epsilon}_{\text{d}}) \left[f(\tilde{\epsilon}_{m}) - 1\right]
   \\
   & =
   \frac{\rho}{V} \int_{-\xi_{0}}^{\xi_{0}} d\tilde{\epsilon} \,
   \frac{f(\tilde{\epsilon}) - f(\tilde{\epsilon}_{\text{d}})}{\tilde{\epsilon}
   - \tilde{\epsilon}_{\text{d}}-iX_{n}}.
   \label{eq:chi^Mat_0_beta-integral}
\end{align}
To arrive at the second expression we solved the integral over $\tau$ and
exploited
\begin{equation}
   f(\tilde{\epsilon}_{\text{d}}) \left[f(\tilde{\epsilon}_{m}) - 1\right]
   \left[e^{(\tilde{\epsilon}_{\text{d}} - \tilde{\epsilon}_{m})\beta} -1\right]
   =
   f(\tilde{\epsilon}_{\text{d}}) - f(\tilde{\epsilon}_{m}).
\end{equation}
Furthermore, we used that the Hartree dressed energies $\tilde{\epsilon}_{m}$
form a band from $-\xi_{0}$ to $\xi_{0}$ with density $\rho,$ as explained above
in Sec.~\ref{subsec:Hartree-dressed-propagators}. In the limit $T\rightarrow0$
holds $f(\tilde{\epsilon}) - f(\tilde{\epsilon}_{\text{d}}) \rightarrow
-\Theta(\tilde{\epsilon})$ such that
\begin{equation}
   \chi_{0,\beta}^{\text{Mat}}(iX_{n}) \xrightarrow{\beta\rightarrow\infty}
   \chi_{0,\infty}^{\text{Mat}}(iX)
   =
   \frac{\rho}{V} \Ln \left(\frac{-\tilde{\epsilon}_{\text{d}}-iX}{\xi_{0}
   -\tilde{\epsilon}_{\text{d}}-iX}\right).
\end{equation}
Here, $\Ln$ denotes the complex logarithm with the branch cut chosen along the
negative real axis. 

We observe that the logarithmic divergence for $X\rightarrow0$ is regularized by
the addend $-\tilde{\epsilon}_{\text{d}} > 0$. An additional regularization
appears in the case of finite temperature, where the step of the numerator
$f(\tilde{\epsilon}) - f(\tilde{\epsilon}_{\text{d}})$ in the integrand in
Eq.~(\ref{eq:chi^Mat_0_beta-integral}) is broadened on a scale of a few $T$. 

In order to keep our analysis simple and comparable to our previous approach
within the real-time zero-temperature formalism~\cite{Diekmann2021}, we focus on
the limit of vanishing temperature.  In Sec.~\ref{subsec:Zero-temperature-limit}
below we describe that this limit can be achieved by the transition from
frequency summations to frequency integrations.  By this transition we do not
only avoid the complication of an additional regularizing scale in form of
temperature but also the complication that an RG flow based on a sharp frequency
cutoff would perform leaps at the discrete finite-temperature Matsubara
frequencies. 

There remains to cope with the regularizing effect of
$\tilde{\epsilon}_{\text{d}}$.  Only after an analytic continuation $i X \mapsto
\nu + i\eta = -\tilde{\epsilon}_{\text{d}} + \Delta\nu + i\eta$ with $\Delta\nu
\ll \xi_{0}$ emerges a bare logarithmic divergence in $\Delta\nu / \xi_{0}$.
For higher-order diagrams, too, the logarithmic divergencies appear only after
analytic continuation to real frequencies $\nu$ close to
$-\tilde{\epsilon}_{\text{d}}$.  If, however, the internal frequency
integrations of a diagram are restricted due to an FRG frequency cutoff, the
expressions for the values of higher-order diagrams become more complicated and
the analytic continuation becomes more challenging.  Then it is difficult to
assess whether certain RG approximations capture the leading logarithms.  This
is considerably more transparent in the case $\tilde{\epsilon}_{\text{d}} =
0^{-}$, in which the logarithmic divergence of
$\chi_{0,\infty}^{\text{Mat}}(iX)$ appears already on the imaginary-frequency
axis for $iX\rightarrow0$.  Then one can see whether certain RG approximations
capture the leading logarithms without analytic continuation to the
real-frequency axis.  Results for the case $\tilde{\epsilon}_{\text{d}}=0^{-}$
are, however, only significant if they can be related to the physically relevant
case of finite $\tilde{\epsilon}_{\text{d}} < 0$.  In
Sec.~\ref{subsec:Setting-eps_d=0} we explain that indeed a direct connection
between the two cases is given for the parquet diagrams containing the
leading-logarithmic contributions.


\subsection{Zero-temperature limit for parquet diagrams}
\label{subsec:Zero-temperature-limit}

Within the framework of the real-time zero-temperature formalism it is well
known that the leading-logarithmic contributions to the particle-hole
susceptibility are given by the parquet diagrams with Hartree dressed lines, in
which each bubble comprises one deep-state and one conduction-state
propagator~\cite{Roulet1969}.  Let us now consider the zero-temperature limit of
the same parquet diagrams within Matsubara formalism.  In general, the correct
zero-temperature limit of any Matsubara diagram is obtained by evaluating the
diagram at finite temperature and taking the limit $\beta\rightarrow\infty$ of
the result.  However, the parquet diagrams under consideration are special in
being skeleton diagrams with Hartree dressed lines. On the lines connecting the
parquet vertices there are no insertions of subdiagrams that represent parts of
the self-energy. Due to the absence of such self-energy type insertions on the
propagator lines, the diagram value in the limit $\beta\rightarrow\infty$ can be
obtained by a different evaluation procedure: In time representation we may
evaluate the diagram directly by using the $\beta\rightarrow\infty$ limit of the
Hartree dressed propagator, namely,
\begin{equation}
   G_{\infty}^{\text{Mat}}(\tau)
   =
   e^{-\tilde{\epsilon}\tau} \left[\Theta(-\tilde{\epsilon}) 
   - \Theta(\tau) \right],
   \label{eq:Limit-prop-tau}
\end{equation}
and by integrating imaginary times over the whole axis. Here, the matrix
$\tilde{\epsilon}$ is given by $\tilde{\epsilon}_{k'k} =
\delta_{k'k}\epsilon_{k}$ and $\tilde{\epsilon}_{\text{d}\text{d}} =
\epsilon_{\text{d}} + g\xi_{0} = \tilde{\epsilon}_{\text{d}}$ and
$\tilde{\epsilon}_{\text{d}k} = 0 = \tilde{\epsilon}_{k\text{d}}$ (compare
Sec~\ref{subsec:Hartree-dressed-propagators}).  In frequency representation we
may use the limit propagator
\begin{equation}
   G_{\infty}^{\text{Mat}}(i\omega)
   =
   \frac{1}{i\omega-\tilde{\epsilon}}
   \label{eq:Limit-prop-omega}
\end{equation}
and integrate over continuous Matsubara frequencies instead of summing over
discrete ones.  For general diagrams, this procedure would miss so-called
``anomalous contributions'' which take care of interaction induced changes in
the occupancies of the levels~\cite{Kohn1960,Luttinger1960}.  Basically, the
approach with limit propagators then fails since a series expansion of
$\Theta(-\tilde{\epsilon})$ in Eq.~(\ref{eq:Limit-prop-tau}) around some
$\tilde{\epsilon} < 0$ does not provide a valid approximation of
$\Theta(-\tilde{\epsilon}_\text{ren})$ at some renormalized
$\tilde{\epsilon}_\text{ren} > 0$. But anomalous contributions arise only in
diagrams with self-energy type insertions on the propagator lines, which we do
not consider here.

Given a diagram $D$ for the particle-hole susceptibility we now can distinguish
the following different ways to evaluate it: as $D_{\beta}^{\text{Mat}}(iX_{n})$
in Matsubara formalism at finite temperature $T = 1 / \beta$; as
$D_{\infty}^{\text{Mat}}(iX)$ resulting from $D_{\beta}^{\text{Mat}}(iX_{n})$ in
the limit $\beta \rightarrow \infty$; as $D_{\text{lp}}^{\text{Mat}}(iX)$
evaluated with the limit propagators from Eq.~(\ref{eq:Limit-prop-tau})
or~(\ref{eq:Limit-prop-omega}); and finally as $D_{\Psi_{0}}(\nu)$ evaluated by
use of the real-time zero-temperature formalism as contribution to the
particle-hole susceptibility in the state $|\Psi_{0}\rangle$.

Consider now any single Matsubara diagram $D_{\text{lp}}^{\text{Mat}}(iX)$ for
the particle-hole susceptibility that is evaluated with Hartree dressed limit
propagators just as described above.  For $\tilde{\epsilon}_{\text{d}}<0$, its
value is indeed just the analytic continuation of the value of the very same
diagram evaluated as $D_{\Psi_{0}}(\nu)$ within the real-time zero-temperature
formalism; we demonstrate this in
Appendix~\ref{sec:App-identification-Matsubara-0T}. Since this way of evaluating
Matsubara diagrams at zero temperature is correct for parquet diagrams with
Hartree dressed lines and since these diagrams are known to contain all
leading-logarithmic contributions to the particle-hole susceptibility in the
real-time zero-temperature formalism, we infer that they contain all these
contributions also in the Matsubara formalism.  The leading-logarithmic
contribution to $\chi^\text{Mat}_\infty(iX)$ can thus be determined by
evaluating parquet diagrams $D_{\text{lp}}^{\text{Mat}}(iX)$ with limit
propagators.


\subsection{Setting $\tilde{\epsilon}_{\text{d}}=0^{-}$}
\label{subsec:Setting-eps_d=0}

In Sec.~\ref{subsec:Regularizing-effects} we explained that it is technically
preferable for the analysis of the leading-logarithmic contributions to work at
zero temperature and vanishing $\tilde{\epsilon}_{\text{d}}$.  Concerning the
temperature, we performed the step to $T=0$ in
Sec.~\ref{subsec:Zero-temperature-limit} where we found that we can use the
limit propagators from Eqs.~(\ref{eq:Limit-prop-tau}) and
(\ref{eq:Limit-prop-omega}) to evaluate the leading-logarithmic parquet
diagrams. Now we focus on $\tilde{\epsilon}_{\text{d}}$ and argue that the
diagram values resulting for the technically desirable choice
$\tilde{\epsilon}_{\text{d}} = 0^{-}$ are directly related to the physically
relevant case of finite $\tilde{\epsilon}_{\text{d}} < 0$.

In Appendix~\ref{sec:App-identification-Matsubara-0T} we show that diagrams
$D_{\text{lp}}^{\text{Mat}}(z)$ for the particle-hole susceptibility
$\chi_\infty^{\text{Mat}}(z)$ evaluated within the Matsubara formalism using
limit propagators are, in the region $\Imag z > 0$ and $\Real z <
-\tilde{\epsilon}_{\text{d}}$, identical to the same diagrams evaluated within
the real-time zero-temperature formalism.  In the special case
$\tilde{\epsilon}_{\text{d}} = 0^{-}$ this region still includes the upper half
of the imaginary $z$-axis. Furthermore, in Ref.~\cite{Diekmann2021} we described
that a Ward identity connects the results for $\chi_{\Psi_{0}}(\nu)$ computed
within the real-time zero-temperature formalism for $\tilde{\epsilon}_{\text{d}}
< 0$ and for $\tilde{\epsilon}_{\text{d}} = 0$ in the form
$\chi_{\Psi_{0}}(\nu,\tilde{\epsilon}_{\text{d}}) =
\chi_{\Psi_{0}}(\nu+\tilde{\epsilon}_{\text{d}},0)$ (see
also~\cite{Roulet1969}). This identity holds diagram by diagram.  It follows for
any parquet diagram $D(\nu,\tilde{\epsilon}_{\text{d}})$ for the susceptibility
with Hartree dressed lines and real-valued $X>0$ and $\nu$ that
\begin{equation}
   \label{eq:restore-epsilon-d}
   D_{\text{lp}}^{\text{\text{Mat}}}(iX,0^{-})\big|_{iX\rightarrow\nu+i\eta}
   =
   D_{\Psi_{0}}(\nu,0)
   =
   D_{\Psi_{0}}(\nu-\tilde{\epsilon}_{\text{d}},\tilde{\epsilon}_{\text{d}}).
\end{equation}
Here, the analytic continuation induced by $iX\rightarrow\nu+i\eta$ requires
formally $\nu<0$.  However, we can directly extend the result to $\nu>0$ because
$D_{\Psi_{0}}(\nu,0)$ allows for an analytic continuation from $\nu<0$ to
$\nu>0$ via the upper half plane.  The same result then holds for the
leading-logarithmic parquet approximation to $\chi$ as sum over such diagrams.
It provides a direct connection between the leading-logarithmic Matsubara
susceptibility computed at $\tilde{\epsilon}_{\text{d}} = 0^{-}$ with limit
propagators and the leading-logarithmic ground-state susceptibility at
physically relevant values of $\tilde{\epsilon}_{\text{d}} < 0$.  Therefore, we
can restrict our Matsubara analysis to the case $\tilde{\epsilon}_{\text{d}} =
0^{-}$, using the deep-state propagator
\begin{equation}
   G_{\infty,\text{d}}^{\text{Mat}}(i\omega)
   =
   \frac{1}{i\omega-0^{-}}.
   \label{eq:Gd-infty-0+}
\end{equation}
 We note that this argument does not rely on whether
$\tilde{\epsilon}_{\text{d}} = 0^{-}$ implies $\epsilon_{\text{d}} <
\epsilon_{\text{d0}}$ or not. 


\section{Leading-logarithmic approximation by one-loop FRG}
\label{sec:Leading-log-approx}

In this section we construct a one-loop Matsubara FRG approximation to the model
at hand which reproduces identically the leading-logarithmic parquet
approximation of Ref.~\cite{Roulet1969}.  We start with an approximation to the
local conduction-electron propagator which we take over from
Refs.~\cite{Lange2015, Kugler2018, Kugler2018a} and which we show to comply with
the leading-logarithmic approximation.  The decisive step is then to realize
that this approximation makes the computational expressions for the Matsubara
diagrams for $\chi^\text{Mat}_\infty(iX)$ coincide completely with those for the
real-time zero-temperature diagrams for $\chi_{\Psi_0}(X)$.  From that point on
we can copy identically all approximation steps that we performed in our
previous real-time zero-temperature approach from Ref.~\cite{Diekmann2021}.


\subsection{Approximation to the local conduction-electron propagator}

In this subsection we make and justify an approximation for the
conduction-electron propagator which was already used in Refs.~\cite{Lange2015,
Kugler2018, Kugler2018a}; there it was motivated cursorily but not justified in
detail.  We show that this approximation does not affect the leading logarithms. 

Consider some parquet diagram $D_{\text{lp}}^{\text{Mat}}(iX)$ within Matsubara
formalism, with lines representing the zero-temperature limit
$G^\text{Mat}_\infty$ of the deep-state propagator or the Hartree dressed
conduction-electron propagator, respectively. Since the value of a bare vertex
is independent of the momenta of the attached conduction-electron lines, all
momentum summations appearing in the diagram are independent of each other.
Therefore, these summations can be performed separately for each
conduction-electron line, leading to the local conduction-electron propagator
\begin{align}
   G_{\infty,\text{c}}^{\text{Mat}}(i\omega)
   &=
   \frac{1}{V} \sum_{k} G_{\infty,k}^{\text{Mat}}(i\omega) 
   \\
   & =
   \frac{\rho}{V}\int_{-\xi_{0}}^{\xi_{0}}
   \frac{d\tilde{\epsilon}}{i\omega-\tilde{\epsilon}}
   \\
   & =
   -2i \frac{\rho}{V} \arctan \frac{\xi_{0}}{\omega}
   \\
   & =
   -i \pi \frac{\rho}{V} \left[\sgn(\omega) - \frac{2}{\pi}
   \arctan\frac{\omega}{\xi_{0}}\right].
   \label{eq:local-cond-el-prop}
\end{align}

We follow Refs.~\cite{Lange2015, Kugler2018, Kugler2018a} in approximating
$G_{\infty,\text{c}}^{\text{Mat}}(i\omega)$ as
\begin{equation}
   G_{\infty,\text{c}}^{\text{Mat}}(i\omega)
   \approx
   -i \pi \frac{\rho}{V} \sgn(\omega) \Theta(\xi_{0}-|\omega|).
   \label{eq:local-cond-el-prop-approx}
\end{equation}
Let us make sure that this approximation does not affect the leading logarithms.
For that purpose we consider one of the bubbles in a leading parquet diagram. We
assume it to be a particle-hole bubble with one deep-state line and one
conduction-electron line; the argument can be applied analogously to the
remaining case of a particle-particle bubble. The value of the particle-hole
bubble is determined by the integral
\begin{equation}
   \int_{-\infty}^{\infty} \frac{d\omega}{2\pi} \, 
   G_{\infty,\text{d}}^{\text{Mat}}(i\omega) 
   G_{\infty,\text{c}}^{\text{Mat}}[i(\omega+X)]
   f(\omega).
   \label{eq:ph-bubble-with-inner-bubbles}
\end{equation}
Here, $X$ denotes the bosonic exchange frequency of the bubble.  The function
$f(\omega)$ results from the more inner bubbles that are contained in the
effective vertices at the two ends of our particle-hole bubble.  $f$ depends in
general on $X$ and on the external frequencies of the bubble; this dependence is
not shown in our notation.  For $|\omega| / \xi_{0} \rightarrow \infty$ and for
$\omega$ approaching certain $\omega_{j}$ which depend on $X$ and on the
external frequencies of our bubble, $f(\omega)$ diverges like some power of a
logarithm.  When combined with the two propagators in
Eq.~(\ref{eq:ph-bubble-with-inner-bubbles}), the factor $f(\omega)$ does not
lead to a divergence of the integral.  In the next two paragraphs we concentrate
on the case $|X| \ll \xi_{0}$ and argue afterwards that this is indeed the
relevant case.

Both propagators in the integrand in Eq.~(\ref{eq:ph-bubble-with-inner-bubbles})
fall off as $1/\omega$ for $|\omega| / \xi_{0} \rightarrow \infty$ so that the
contribution from large frequencies $|\omega| \gtrsim \xi_{0}$ to the integral
is $\mathcal{O}(1)$.  Therefore, we can neglect the contribution from this
frequency region.  The lower bound for the negligible region of $|\omega|$ is
somewhat arbitrary; the only condition is that $\omega$ with $|\omega| \ll
\xi_{0}$ are not neglected.  In particular the negligible region can be chosen
as $|X+\omega| \ge \xi_{0}$ which justifies the step function in
Eq.~(\ref{eq:local-cond-el-prop-approx}).  We note that such a cutoff in the
approximated propagator is required:  An insertion of the propagator from
Eq.~(\ref{eq:local-cond-el-prop-approx}) without the step function into the
integral in Eq.~(\ref{eq:ph-bubble-with-inner-bubbles}) would cause the
contribution from large frequencies $|\omega| \gtrsim \xi_{0}$ to diverge.

The leading-logarithmic contributions to the integral in
Eq.~(\ref{eq:ph-bubble-with-inner-bubbles}) originate from the combination of
the factor $G_{\infty,\text{d}}^{\text{Mat}}(i\omega) = 1/(i\omega-0^{-})$ with
the discontinuous function $\sgn(\omega+X)$ contained in
$G_{\infty,\text{c}}^{\text{Mat}}[i(\omega+X)]$ in
Eq.~(\ref{eq:local-cond-el-prop}).  In fact, this mechanism of how the
logarithmic divergencies arise is the same as in the zero-temperature formalism
(cf. Sec.~III\,D of Ref.~\cite{Diekmann2021}). The important frequency range is
given by $|X| \le |\omega| \ll \xi_{0}$, where we can approximate
\begin{equation}
   G_{\infty,\text{c}}^{\text{Mat}}[i(\omega+X)]
   \approx
   -i \pi \frac{\rho}{V} \sgn(\omega+X).
\end{equation}
A straight computation shows that the next correction term to $G_{\infty,
\text{c}}^{\text{Mat}}[i(\omega+X)]$, which is of order
$(X+\omega)/\xi_{0}$, yields only a subleading contribution.

We still need to justify that we restricted our considerations to the case
$|X|\ll\xi_{0}$.  The argument will first be given for the outermost bubbles in
a diagram for the particle-hole susceptibility.  Then it can be replicated
iteratively for the more and more inner tiers.  To start with, the
leading-logarithmic approximation for the susceptibility
$\chi_\infty^{\text{Mat}}(iX)$ is good only for $|X|\ll\xi_{0}$, for which the
logarithms become large.  This corresponds to the fact that after an analytic
continuation $iX \rightarrow \nu + i\eta$ we are interested in the behavior of
$\chi(\nu)$ near threshold.  Now, the bosonic exchange frequency $X_{1}$ of the
outermost bubbles in a diagram $D_{\text{lp}}^{\text{Mat}}(iX)$ contributing to
$\chi_\infty^{\text{Mat}}(iX)$ is equal to $X$ and thus $|X_1| \ll \xi_{0}$.
Concerning the next inner bubbles, we consider some ``crossed''
particle-particle bubble which may be contained in the effective vertex between
two outermost particle-hole bubbles.  If we call the integration frequencies of
these outermost particle-hole bubbles $\omega_{1}$ and $\omega'_{1}$, then the
crossed particle-particle bubble has the bosonic total frequency $\Omega_{2} =
\omega_{1} + \omega'_{1} + X_{1}$. We discussed above that the important range
of frequency integration in the particle-hole bubbles is given by $|X_{1}| \le
|\omega_{1}^{(\prime)}| \ll \xi_{0}$. Hence, $|\Omega_{2}| \ll \xi_{0}$ for the
important contributions.  In turn, the leading contributions to the value of the
particle-particle bubble arise from a frequency integration over $\omega_{2}$
with $|\Omega_{2}| \le |\omega_{2}| \ll \xi_{0}$; this can be shown on the
analogy of the above discussion for the particle-hole bubble.  If now the
effective vertex between two particle-particle bubbles with integration
frequencies $\omega_{2}$ and $\omega_{2}'$ contains a more inner, crossed
particle-hole bubble, then the latter one is characterized by the bosonic
exchange frequency $X_{3} = \Omega_{2} - \omega_{2} - \omega'_{2}$ which again
satisfies $|X_{3}| \ll \xi_{0}$.  This argument can be repeated on and on to
show that the leading contributions result only from those internal integration
frequencies for which the natural bosonic frequencies $X_i, \Omega_i$ of all
inner bubbles are much smaller that $\xi_{0}$.


\subsection{Identity between leading-logarithmic contributions in real and
imaginary time} \label{sec:identity-leading-log-real-imag}

In this subsection we study the consequences of
approximation~(\ref{eq:local-cond-el-prop-approx}).  We find that the
approximate expressions that we obtain for the Matsubara diagrams of the
particle-hole susceptibility on the imaginary-frequency axis are plainly
identical (without analytic continuation) to the ones known from the real-time
zero-temperature approach on the real axis. This makes it possible to use just
the identical FRG approximation steps as used in the real-time case in
Ref.~\cite{Diekmann2021}.

First we observe that the approximated local conduction-electron propagator from
Eq.~(\ref{eq:local-cond-el-prop-approx}) coincides with the imaginary part of
the local conduction-electron propagator $G_{\text{c}}(\omega)$ from the
real-time zero-temperature formalism,
\begin{equation}
   G_{\text{c}}(\omega)
   =
   \frac{\rho}{V} \left[ \ln \frac{|\xi_{0}+\omega|}{|\xi_{0}-\omega|}
   - i \pi \sgn(\omega) \Theta(\xi_{0} - |\omega|) \right].
\end{equation}
The imaginary part of this propagator is indeed the only part that was retained
during the leading-logarithmic calculations in the real-time zero-temperature
formalism (see Refs.~\cite{Roulet1969, Diekmann2021}). 

Furthermore, the deep-state propagator in Eq.~(\ref{eq:Gd-infty-0+}) is, up to a
factor $(-i)$, identical to the deep-state propagator used in the treatment of
the model with the real-time zero-temperature formalism (see
Refs.~\cite{Roulet1969, Diekmann2021}).  We note that the particular choice
$\tilde \epsilon_\text{d}=0^-$ is responsible for this match between $1/(i\omega
- \tilde \epsilon_\text{d})$ and $(-i)/(\omega - \tilde
\epsilon_\text{d}-i0^+)$. 

The diagram rules for the susceptibility in Matsubara formalism differ from
those in the real-time zero-temperature formalism by a global factor $i$ and by
factors $i$ for each vertex.  All these factors $i$ together precisely
compensate the factors $(-i)$ in the deep-state propagators.  As a consequence,
the resulting expressions for the Matsubara diagrams
$D_{\text{lp}}^{\text{Mat}}(iX)$ computed with approximated local
conduction-electron propagators are literally identical to those for the
diagrams for $\chi_{\Psi_{0}}(X)$ in the real-time zero-temperature formalism.  

This conformance needs to be distinguished from the analytic continuation
between $\chi_\infty^{\text{Mat}}$ and $\chi_{\Psi_{0}}$:  Without the
approximation~(\ref{eq:local-cond-el-prop-approx}) for the Matsubara
conduction-electron propagator, the diagrams $D_{\text{lp}}^{\text{Mat}}(z)$ for
$\chi_\infty^{\text{Mat}}(z)$ are identical in value to those for
$\chi_{\Psi_{0}}(z)$ in the upper left quadrant of $z$ (see
Appendix~\ref{sec:App-identification-Matsubara-0T}).  In particular the diagrams
$D_{\text{lp}}^{\text{Mat}}(iX)$  have the same values as the diagrams for
$\chi_{\Psi_{0}}(iX)$, not as those for $\chi_{\Psi_{0}}(X)$ as found above.
However, in the leading-logarithmic approximation the diagram values for
$\chi_{\Psi_{0}}(iX)$ and for $\chi_{\Psi_{0}}(X)$ can both be reduced to the
same factors of the form $\ln|X|/\xi_{0}$ and are thus identical.  This can be
understood from the result
\begin{equation} 
   \label{eq:chi-Roulet-final}
   \chi_{\Psi_0}(z) 
   =
   \frac{\rho}{2g} 
   \left[1 - \left(-\frac{\xi_0}{z+\tilde\epsilon_\text{d}}\right)^{2g}\right],
\end{equation}
which was constructed in Ref.~\cite{Roulet1969} by imposing the desired analytic
properties onto the leading-logarithmic approximation.  According to this
result, the leading contribution to $\chi_{\Psi_0}(X)$ and that to
$\chi_{\Psi_0}(iX)$ at $\tilde \epsilon_\text{d}=0^-$ read both as $\rho \left[1
- (\xi_0/|X|)^{2g}\right]/(2g)$.  

For the simple example of the noninteracting particle-hole susceptibility, the
peculiar approximate identity between $\chi_\infty^{\text{Mat}}(iX)$ and
$\chi_{\Psi_{0}}(X)$ is traced back to the pole structure of the integrand in
the complex plane in Appendix~\ref{sec:App-illustrate-identity}.


\subsection{Adopting the steps from the real-time zero-temperature approach}

Given the identity between the approximate expressions for the diagrams
$D_{\text{lp}}^{\text{Mat}}(iX)$ for $\chi_\infty^{\text{Mat}}(iX)$ and those
for $\chi_{\Psi_{0}}(X)$, it is straightforward to construct a one-loop
Matsubara FRG approximation that captures all leading logarithms: We can copy
one by one the steps from the zero-temperature FRG approach of
Ref.~\cite{Diekmann2021}.  Here we summarize these steps only very briefly; for
details see Sec.~V of Ref.~\cite{Diekmann2021}.

First, we introduce a sharp frequency cutoff into the principal-value part of
the Hartree dressed deep-state propagator:
\begin{equation}
   \label{eq:cut-off}
   G_{\infty, \text{d}}^{\text{Mat}, \lambda}(i\omega) 
   =
   -i \Theta(|\omega|-\lambda) \frac{1}{\omega} + \pi \delta(\omega).
\end{equation}
Then the conduction-state propagator at the initial flow parameter
$\lambda_\text{ini} \rightarrow \infty$ proves to be Hartree dressed.  We
approximate this propagator by Eq.~(\ref{eq:local-cond-el-prop-approx}).  In
leading-logarithmic order the initial value of the (one-particle irreducible)
two-particle vertex function is given by the bare interaction while all higher
vertex functions vanish.

Next we neglect the flow of the self-energy, of the three-particle vertex
function and of the two-particle vertex function with four deep-state indices.
In the flow equations for the two-particle vertex function with two deep-state
and two conduction-state indices we perform the logarithmic approximation by
setting, e.g., $|\omega + \lambda - \Omega| \approx \max\{|\omega|,\lambda\}$ in
the frequency arguments of the vertex functions.  Furthermore, we approximate
the start of the flow by setting, e.g., $\Theta(\xi_0 - |\Omega+\lambda|)
\approx \Theta(\xi_0 - \lambda)$. 

As shown in Ref.~\cite{Diekmann2021}, we then obtain the identical integral
equations for the two-particle vertex function as in Ref.~\cite{Roulet1969}.  We
copy the steps for its solution and for the computation of the susceptibility
and obtain
\begin{equation}
   \label{eq:leading-log-result-imag-axis}
   \chi^\text{Mat}_\infty(iX) 
   =
   \frac{\rho}{2g} \left[1 - \left(\frac{\xi_0}{|X|}\right)^{2g} \right].
\end{equation}
We identify this function with the analytic continuation of the retarded
ground-state susceptibility, $\chi^\text{Mat}_\infty(iX) =
\chi^\text{ret}_{\Psi_0}(iX) = \chi_{\Psi_0}(iX)$ (see
Sec.~\ref{subsec:Linear-response-rate}).  Then the only difference compared to
Refs.~\cite{Roulet1969,Diekmann2021} is that we obtain the
result~(\ref{eq:leading-log-result-imag-axis}) for $\chi_{\Psi_0}(iX)$ instead
of $\chi_{\Psi_0}(X)$.  This difference is of no importance:  We can reconstruct
the imaginary part of $\chi_{\Psi_0}(z)$ in the same way as
Ref.~\cite{Roulet1969}, namely, such that the branchcut of the function is
located on the positive semiaxis, as required.  We end up again with
Eq.~(\ref{eq:chi-Roulet-final}).


\section{Comparison to Ref.~\cite{Kugler2018a}}
\label{sec:Comparison-to-Kugler}

In Ref.~\cite{Kugler2018a} the same model of x-ray absorption is investigated
with different one-loop and multiloop FRG approximations.
Reference~\cite{Kugler2018a} presents data which show that, compared to one-loop
FRG, multiloop iterations improve the agreement of the FRG results with those of
a numerical solution of the parquet approximation.  We should now clarify how
this relates to our construction of a one-loop Matsubara FRG scheme that
reproduces identically the leading-logarithmic parquet approximation of Roulet
\emph{et al.}~\cite{Roulet1969}. There are two possible explanations for the
improved agreement between multiloop FRG and the numerical parquet results which
is observed in Ref.~\cite{Kugler2018a}.  One is that the one-loop FRG
approximations of Ref.~\cite{Kugler2018a} are not constructed properly and miss
certain leading contributions.  The other is that the numerical improvements due
to multiloop FRG in Ref.~\cite{Kugler2018a} are subleading and thus beyond the
controlled regime of the parquet approximation. 

The second possibility is actually very plausible.  We note that due to several
approximations in the analytic evaluation, even the leading-logarithmic parquet
result of Roulet \emph{et al.}~differs subleadingly from the numerically
evaluated sum of the parquet diagrams.  On the subleading level both
approximations are uncontrolled and none is thus preferable \emph{a priori}.
Figure~4(d) of Ref.~\cite{Kugler2018a} displays the particle-hole susceptibility
at zero imaginary frequency as function of the interaction strength.  According
to this figure, the agreement between the leading-logarithmic result of Roulet
\emph{et al.}~and the numerical sum of the parquet diagrams is fine for
interactions up to roughly $g=0.2$, while larger values of $g$ (called $u$ in
that reference) lead to sizable differences.  The figure caption attributes the
differences to subleading contributions; they are thus in the uncontrolled
regime.  In this context we note that while the choice $2g-g^2$ for the exponent
of the guide-to-the-eye lines in this figure represents the correct function in
second order (compare Ref.~\cite{Nozieres1969a}), it is uncontrolled in the
parquet approximation; the correction $-g^2$ was even renounced in the improved
self-consistent treatment of Ref.~\cite{Nozieres1969}.  If we concentrate on
$g=0.28$, we find for $|\chi^\text{Mat}(iX\!=\!0)|$ from Fig.~4 of
Ref.~\cite{Kugler2018a} approximately $7 \rho$ for the numerical parquet
computation and approximately $9 \rho$ for the leading-logarithmic result of
Roulet \emph{et al.}  The difference of about $2 \rho$ between these values
highlights the large influence of subleading contributions.  From the ordinate
intercepts in Fig.~5 of Ref.~\cite{Kugler2018a} we infer that the one-loop FRG
results for $|\chi^\text{Mat}(iX\!=\!0)|$ depend on the choice of regulator and
are roughly in the range from $8.5 \rho$ to $11 \rho$.  This differs again up to
$2\rho$ from the $9\rho$ corresponding to the solution of Roulet \emph{et al.}
It is hence reasonable to assume that all approximations with different loop
orders studied in Ref.~\cite{Kugler2018a} differ only subleadingly. 

In spite of this finding we have the impression that the regulators chosen in
Ref.~\cite{Kugler2018a} are not optimal for a one-loop FRG.  They suffer from
the drawback that the dependence of the regulator dressed deep-state propagators
on $\omega$ and $\epsilon_\text{d}$ is not restricted to a function of
$i\omega-\epsilon_\text{d}$ alone.  In fact, the regulator-free bare deep-state
propagator $1/(i\omega-\epsilon_\text{d})$ is a function of
$i\omega-\epsilon_\text{d}$ only.  This property ensures that the particle-hole
susceptiblity $\chi^\text{Mat}(z)$ is a function of $z+\epsilon_\text{d}$
(compare Sec.~III\,C of Ref.~\cite{Diekmann2021}).  This Ward identity related
to particle-number conservation and time translational invariance is satisfied
not only by the exact result, but also by any single diagram, by the sum of the
parquet diagrams and by the leading-logarithmic result from
Eq.~(\ref{eq:chi-Roulet-final}). Correspondingly, the very logarithmic
divergencies appearing in perturbation theory are of the form
$\Ln[(z+\epsilon_\text{d})/\xi_0]$ and are thus divergencies in
$z+\epsilon_\text{d}$. The different regulators investigated in
Ref.~\cite{Kugler2018a} destroy this property.  Among those is, for example, a
sharp frequency cutoff in the deep-state propagator,
\begin{equation}
   \label{eq:cut-off-Kugler}
   G_\text{d}^\lambda(i\omega_n)
   =
   \Theta(|\omega_n| - \lambda) 
   \frac{1}{i\omega_n - \epsilon_\text{d}},   
\end{equation}
with $\lambda$ denoting the flow parameter.  Here, the cutoff prefactor
$\Theta(|\omega_n| - \lambda)$ is not a function of $(i\omega_n -
\epsilon_\text{d})$.  The resulting defect of the one-loop approximation is
healed by the multiloop iterations since they make the particle-hole
susceptibility converge to the sum of the parquet diagrams. The Ward identity is
then satisfied again so that the nature of the divergencies can be captured.

We note in passing that our one-loop flow induced by the cutoff from
Eq.~(\ref{eq:cut-off}) is not affected by this problem.  As we set $\tilde
\epsilon_\text{d}=0^-$, the result of the flow depends on $z$ alone.  After the
flow, the use of Eq.~(\ref{eq:restore-epsilon-d}) allows for the transition from
$z$ to $z+\tilde \epsilon_\text{d}$, which then appears correctly in
Eq.~(\ref{eq:chi-Roulet-final}).

In summary, Ref.~\cite{Kugler2018a} demonstrates with specific examples that the
multiloop FRG scheme allows to construct the sum of the parquet diagrams even by
use of cutoffs whose suitability for the model was not inspected in detail. The
resulting numerical parquet approximation is then automatically guaranteed to
contain the leading-logarithmic contributions.  In the present paper we
demonstrated how the leading-logarithmic parquet approximation can also be
obtained from a well-constructed one-loop FRG flow. This shows that no general
superiority of multiloop flows over one-loop flows is given for the model at
hand.


\section{Conclusion}
\label{sec:Conclusion}

The leading-logarithmic parquet approximation for certain zero- and
one-dimensional condensed-matter problems is a prominent example of a controlled
approximation in a regime not accessible by perturbation theory.  In a previous
paper~\cite{Diekmann2021} we revealed that a suitably constructed, one-loop
truncated fermionic FRG approximation merges with the leading-logarithmic
parquet approximation of Roulet \emph{et al.}~for x-ray absorption in
metals~\cite{Roulet1969}.  This highlights the capability of the fermionic FRG
to generate controlled nonperturbative approximations, in contrast to the
perturbatively or \emph{a posteriori} justified truncations which are currently
widely used~\cite{Metzner2012,Dupuis2021} in fermionic condensed-matter FRG
studies.  The potential of the fermionic FRG to provide controlled
approximations should now be systematically developed.  The first important step
is to transfer the construction of a leading-logarithmic one-loop FRG from the
zero-temperature formalism used in Ref.~\cite{Diekmann2021} to the Matsubara
formalism, which is most broadly used for applications of the FRG to
condensed-matter
problems~\cite{Metzner2012,Kopietz2010,Platt2013,Salmhofer2019,Dupuis2021}. This
is what we achieved in this paper.

We demonstrated how to construct a one-loop FRG approximation within Matsubara
formalism that leads to the known leading-logarithmic approximation for the
absorption of x rays in metals.  Our approach is founded on the fact that the
leading approximation for the particle-hole susceptibility $\chi_{\Psi_0}(z)$ is
identical on the real and on the imaginary $z$ axis when $\tilde
\epsilon_\text{d}$ is set to zero [see
Eq.~(\ref{eq:leading-log-result-imag-axis})].  Due to this property it was
possible to find an approximate formulation of the Matsubara perturbation theory
at $T=0$ which leads to imaginary-frequency expressions that are identical to
the real-frequency expressions from the zero-temperature ground-state formalism.
At that stage we could copy all further steps from Ref.~\cite{Diekmann2021}.

We thus substantiated that, also within Matsubara formalism, leading-logarithmic
approximations can be achieved by one-loop FRG approximations.  We expect that a
generalization of our one-loop flow to $T>0$ is possible in a natural way by
using discrete Matsubara frequencies and then, potentially, resorting to a
numerical solution.  This would allow to study how the temperature-induced
regularization of the divergencies influences observables.  For $T\rightarrow 0$
the results of such an approach are expected to converge to those of the
continuous-frequency approach presented in this paper. Furthermore, we expect
that our analysis can be transferred to models whose structure of logarithmic
divergencies in perturbation theory resembles that of the x-ray absorption
model, in particular to the Kondo model and to the Fermi gas model of
one-dimensional conductors.

We pointed out that the multiloop corrections to the one-loop results in
Ref.~\cite{Kugler2018a} are of a size corresponding to subleading contributions.
This is, however, not important for the main message of that reference, which is
that the multiloop scheme ensures the numerical summation of the complete
parquet diagrams for any choice of cutoff.  The approximation resulting from
that summation is known to satisfy certain symmetries, sum rules, and
conservation laws~\cite{Bickers1991,Vilk1997,Kugler2018c}.  This was beneficial
for multiloop FRG studies of two-dimensional systems of correlated
electrons~\cite{Tagliavini2019,Hille2020a,Hille2020} and three-dimensional
interacting quantum spin systems~\cite{Kiese2022,Ritter2022}. But concerning the
particular class of systems represented by the x-ray absorption model with its
plain structure of logarithmic divergencies, we established that the desirable
leading-logarithmic approximation can be achieved even analytically by a
suitably constructed one-loop FRG.

A broad field of future investigations opens up on the basis of our present
study.  From a methodological point of view, two extensions are highly
desirable: on the one hand, that to a consistent handling of subleading
contributions, on the other hand, that to the nonequilibrium Keldysh formalism.
From an applications point of view, our scheme awaits to be adapted to the study
of diverse zero- and one-dimensional quantum-dot systems at low temperatures in
and out of equilibrium.  A long term goal is the construction of an efficient
and analytically transparent impurity solver whose range of applicability
extends from the perturbative to the Kondo regime.


\acknowledgments

We thank V.~Meden for helpful comments on the manuscript. This work was
supported by the Deutsche Forschungsgemeinschaft via Grant No.~RTG~1995.



\appendix

\section{Details on the state $\nket{\bar{\Psi}_{0}}$
\label{sec:App-Details-on-ground-state}}

In Sec.~\ref{subsec:Ground-state} we described that the ground state of the
system filled with $(N+1)$ particles is either the state $\ket{\Psi_{0}}$ in
which the deep level and the $N$ states in the lower half of the conduction band
of momentum states are occupied, or the state $\nket{\bar{\Psi}_{0}}$ in which a
bound state and $N$ states in the lower part of the band of scattering states
are occupied.  The bound state and the scattering states are single-particle
eigenstates of the perturbed Hamiltonian
\begin{equation}
   \bar{H}_{0}
   =
   \sum_{k,k'} \left(\delta_{kk'}\epsilon_{k} - \frac{U}{V}\right) 
   a_{k'}^{\dagger} a_{k}.
\end{equation}
 Here we analyze these eigenstates of $\bar{H}_{0}$ in more detail.

\begin{figure}
   \begin{center}
      \includegraphics[width=\columnwidth]{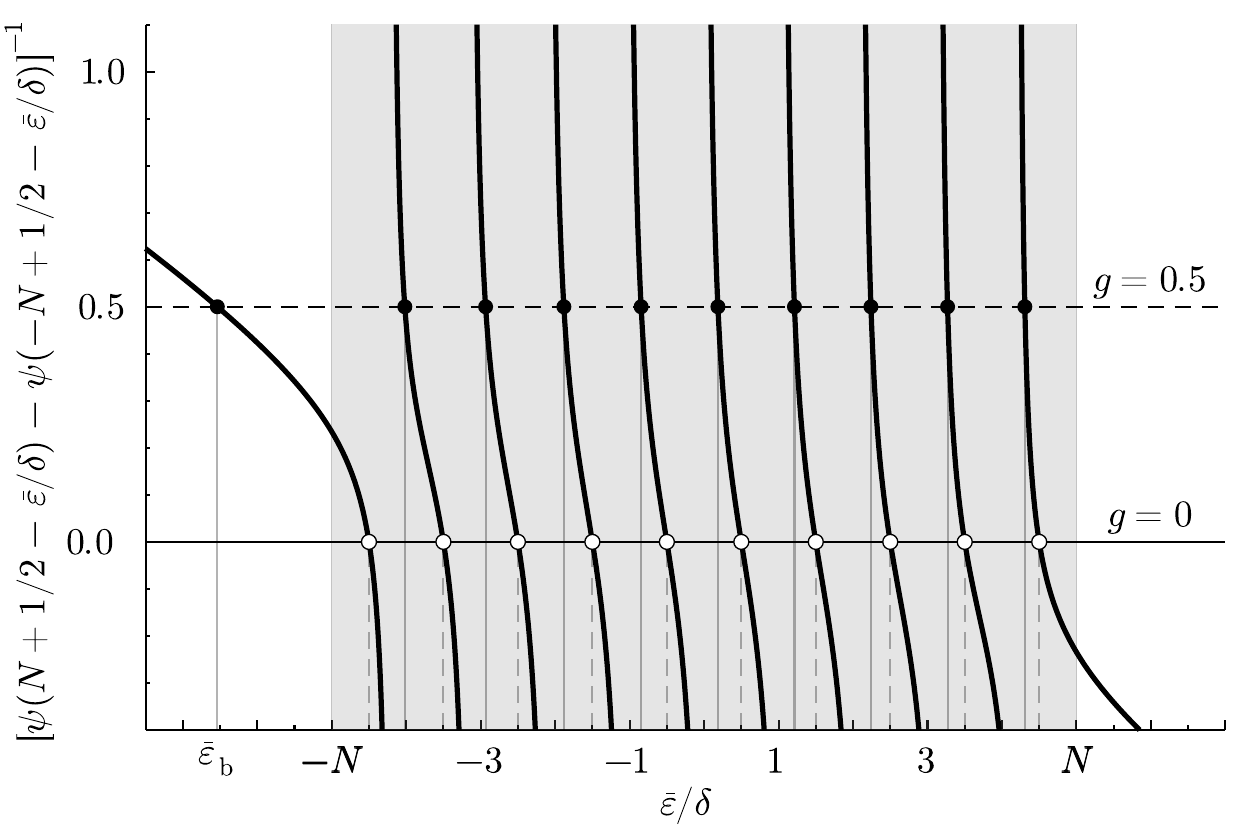}
   \end{center}
   \caption{Graphical solution of the implicit
      equation~(\ref{eq:perturbed-energies-implicit-psi}) for the case $N=5$.
      (The value $N=5$ is sufficiently small to make the details visible and
      sufficiently large to convey the impression of a band of states. Realistic
      values of $N$ would be much larger.) The empty circles mark singularities
      of the denominator $\psi\left(N + \tfrac{1}{2} -
      \bar{\epsilon}_{n}/\delta\right) - \psi\left(-N + \tfrac{1}{2} -
      \bar{\epsilon}_{n}/\delta\right)$, corresponding to the positions of the
      unperturbed energies $\epsilon_{n} = \left(n-N-\frac{1}{2}\right)\delta$
      at $g=0$.  The gray region visualizes the corresponding energy window of
      the unperturbed conduction band.  The black dots show the solutions of the
      implicit equation for $g=0.5$ and thus mark the positions of the
      corresponding perturbed energies $\bar{\epsilon}_{n}$.}
   \label{fig:Graphical-solution-of-impl-eq}
\end{figure}

The single-particle energies of $\bar{H}_{0}$ are the eigenvalues
$\bar{\epsilon}_{n}$ of the matrix with components $\bar{\epsilon}_{kk'} =
\delta_{kk'} \epsilon_{k} - U/V$.  First order perturbation theory yields
\begin{equation}
   \bar{\epsilon}_{n}
   =
   \epsilon_{n} - \frac{g}{\rho} + \mathcal{O}\left(g^{2}\right), \qquad 
   n=1,\dots,2N
   \label{eq:tilde-epsilon-PT}
\end{equation}
with $\epsilon_n \equiv \epsilon_{k_n}$ denoting the unperturbed energies and
with $g = \rho U/V$.  In order to analyze $\bar{\epsilon}_{n}$ beyond
perturbation theory we introduce a discrete model for the band of unperturbed
momentum states: We set $\epsilon_{n} = \left(n-N-\frac{1}{2}\right)\delta,$ $n
= 1,\dots, 2N$, with a constant level spacing $\delta > 0$.  If $\delta$ is much
smaller than all other energies of interest, this discrete band can be treated
as a continuous conduction band with half-width $\xi_{0} = N\delta$ and density
of states $\rho = 1/\delta$.  For the discrete model, a straightforward
computation of the roots $\bar{\epsilon}_{n}$ of the characteristic polynomial
of $\bar{H}_{0}$ in the single-particle sector leads to the implicit equation
\begin{align}
   g 
   & =
   \left( 
      \sum_{j=0}^{2N-1} \frac{1}{j-N+\frac{1}{2} - \bar{\epsilon}_{n}/\delta}
   \right)^{-1}
   \label{eq:perturbed-energies-implicit}
   \\
   & =
   \left[ 
      \psi\left(N+\tfrac{1}{2}-\bar{\epsilon}_{n}/\delta\right) 
      - \psi\left(-N+\tfrac{1}{2}-\bar{\epsilon}_{n}/\delta\right) 
   \right]^{-1},
   \label{eq:perturbed-energies-implicit-psi}
\end{align}
with $g = U/(V\delta) \neq 0$, and with the digamma function $\psi$ used in the
second line to express the sum appearing in the first line; for details on the
digamma function see Chap.~5 of Ref.~\cite{DLMF2023}.  This implicit equation
has $2N$ solutions $\bar{\epsilon}_{n}, n = 1,\dots,2N$.  A graphical solution
makes evident that
\begin{equation}
   \bar{\epsilon}_{1} < \epsilon_{1} 
   \qquad\text{and}\qquad
   \epsilon_{n-1} < \bar{\epsilon}_{n} < \epsilon_{n},
   \quad n=2, \dots, 2N
   \label{eq:nested-energies}
\end{equation}
(see Fig.~\ref{fig:Graphical-solution-of-impl-eq}).

Of particular interest is the energy $\bar{\epsilon}_{\text{b}} :=
\bar{\epsilon}_{1}$ of a bound state that results from the localized attractive
potential generated by the deep hole.  An approximation for
$\bar{\epsilon}_{\text{b}}$ that applies to those values of $g$ which are
relevant for our paper can be derived from $\psi(x) \approx \ln x$ for $x \gg
1$.  Under the assumption $-N + \frac{1}{2} - \bar{\epsilon}_{\text{b}}/\delta
\gg 1$ this approximation can be applied to
Eq.~(\ref{eq:perturbed-energies-implicit-psi}), yielding
\begin{equation}
   \frac{1}{g}
   \approx
   \ln\frac{N + \frac{1}{2} - \bar{\epsilon}_{\text{b}}/\delta}{-N 
   + \frac{1}{2} - \bar{\epsilon}_{\text{b}}/\delta}
\end{equation}
with the solution
\begin{equation}
   \bar{\epsilon}_{\text{b}}
   \approx
   \left(-N + \tfrac{1}{2}\right) \delta - \frac{2N\delta}{e^{1/g} - 1}
   \approx
   -\xi_{0}\left(1 + 2e^{-1/g}\right).
   \label{eq:bound-state-energy-approx}
\end{equation}
The last approximation step in Eq.~(\ref{eq:bound-state-energy-approx}) requires
additionally $g \ll 1$.  In total, the above approximation for
$\bar{\epsilon}_{\text{b}}$ is good for $1 \gg g \gg 1/\ln(2N) =
1/\ln(2\xi_{0}\rho)$.

In the $(N+1)$-particle state $\nket{\bar{\Psi}_{0}}$, the bound state and the
states $\bar{k}_{2}, \dots, \bar{k}_{N+1}$ are occupied, while the states
$\bar{k}_{N+2}, \dots, \bar{k}_{2N}$ are empty.  The energy of
$\nket{\bar{\Psi}_{0}}$ is thus
\begin{equation}
   \bar{E}_{0}
   =
   \bar{\epsilon}_{\text{b}}+\bar{E}_{\text{cb}},
\end{equation}
where
\begin{equation}
   \bar{E}_{\text{cb}}
   =
   \sum_{n=2}^{N+1}\bar{\epsilon}_{n}
\end{equation}
is the energy of the half-filled band of scattering states.  From the
inequality~(\ref{eq:nested-energies}) follows
\begin{equation}
   E_{\text{cb}} < \bar{E}_{\text{cb}} < E_{\text{cb}}+\xi_{0}.
   \label{eq:E_cb-E_tilde_cb-nesting-1}
\end{equation}
The difference in energy of the half-filled band of scattering states and the
half-filled band of plane-wave states, $\bar{E}_{\text{cb}} - E_{\text{cb}}$, is
thus less than the single-particle energy $\xi_{0}$.


\section{Linear response rate \label{sec:App-linear-response-rate}}

In Sec.~\ref{subsec:Linear-response-rate} we reported that the mean occupancy of
the deep level changes under the influence of the x-ray field as
\begin{equation}
   \frac{d}{dt}\langle n_{\text{d}}\rangle(t)
   \xrightarrow{t_{0}\rightarrow-\infty}
   2|W|^{2} \Imag \chi^{\text{ret}}(\nu)
\end{equation}
in linear response.  Here we give some details on the derivation.

The rate of change of the mean deep-state occupancy is
\begin{align}
   \frac{d}{dt}\left\langle n_{\text{d}}\right\rangle (t) 
   & =
   \frac{d}{dt}\left\langle U_{\text{tot}}(t_{0},t) n_{\text{d}}
   U_{\text{tot}}(t,t_{0})\right\rangle 
   \\
   & =
   i \left\langle U_{\text{tot}}(t_{0},t) [H_{X}(t), n_{\text{d}}]
   U_{\text{tot}}(t,t_{0})\right\rangle 
   \\
   & =
   i\left\langle U_{\text{I}}(t_{0},t) [H_{X}, n_{\text{d}}](t)_{H}
   U_{\text{I}}(t,t_{0})\right\rangle.
\end{align}
Here, $\left\langle \cdot\right\rangle =\Tr(\varrho\,\cdot)$ denotes the
expectation value given the density operator $\varrho$.  Furthermore,
$U_{\text{tot}}$ denotes the time-evolution operator under the total Hamiltonian
$H_{\text{tot}}(t) = H + H_{X}(t)$, and $U_{I}(t',t) = U_{H}(0,t')
U_{\text{tot}}(t',t) U_{H}(t,0)$ denotes the interaction-picture time-evolution
operator, with $U_{H}$ referring to time evolution under $H$ only. Furthermore, 
\begin{equation}
   [H_{X}, n_{\text{d}}](t)_{H} 
   =
   U_{H}(0,t) [H_{X}(t), n_{\text{d}}] U_{H}(t,0)
\end{equation}
is the commutator in the interaction picture.  By use of the lowest-order
approximation
\begin{equation}
   U_{I}(t,t_{0}) 
   =
   1 - i\int_{t_{0}}^{t} dt_{1}\, H_{X}(t_{1})_{H}
\end{equation}
we obtain
\begin{multline}
   \frac{d}{dt}\left\langle n_{\text{d}}\right\rangle (t)
   =
   i\left\langle [H_{X}, n_{\text{d}}](t)_{H}\right\rangle 
   \\
   + \int_{t_{0}}^{t}dt_1 \left\langle \big[[H_{X}, n_{\text{d}}](t)_{H}, 
   H_{X}(t_1)_{H}\big]\right\rangle.
\end{multline}
The first addend on the right-hand side vanishes since the density operator is
supposed to commute with $H$ and $n_\text{d}$.  For the second addend we employ
\begin{multline}
   \left\langle \big[[H_{X}, n_{\text{d}}](t)_{H}, H_{X}(t_{1})_{H}\big]
   \right\rangle
   =
   -2 |W|^{2} \Real e^{i\nu(t-t_{1})} 
   \\
   \times\left\langle \left[A(t-t_{1})_{H},
   A^{\dagger}\right]\right\rangle 
\end{multline}
to obtain
\begin{align}
   \frac{d}{dt}\langle n_{\text{d}}\rangle(t) 
   & =
   -2 |W|^{2} \Real \int_{0}^{t-t_{0}}dt'\, e^{i\nu t'} \left\langle
   \left[A(t')_{H},A^{\dagger}\right]\right\rangle 
   \\ 
   &\xrightarrow{t_{0}\rightarrow-\infty}
   2 |W|^{2} \Imag \chi^{\text{ret}}(\nu),
\end{align}
as indicated in Sec.~\ref{subsec:Linear-response-rate}.


\section{Analytic behavior of exciton propagators
\label{sec:App-analytic-behaviour}}

In Sec.~\ref{subsec:Linear-response-rate} we reported in which regions the
retarded and Matsubara exciton propagator are analytic and where they coincide.
In this part of the appendix we derive these statements from the Lehmann
representation of the propagators. 

By use of the Lehmann representation it is straightforward to show that
$\chi^{\text{ret}}(z)$ is analytic in the open upper half plane of $z$.  For
that purpose, let $\ket{m}$ denote the states of an orthonormal basis of common
eigenstates of $H$ and $\varrho$, with $H\ket{m} = E_{m}\ket{m}$ and
$\varrho\ket{m} = \varrho_{m}\ket{m}$. Then
\begin{align}
   \chi^{\text{ret}}(z) 
   & =
   -i\int_{0}^{\infty}dt\, e^{izt} \Tr\varrho \left[A(t)_{H} A^{\dagger} - 
   A^{\dagger}A(t)_{H}\right]
   \\
   & =
   -i \sum_{m,l} \int_{0}^{\infty}dt\, e^{i(z+E_{m} - E_{l})t}(\varrho_{m}
   - \varrho_{l}) A_{ml} A_{lm}^{\dagger}
   \label{eq:chi-ret-Lehmann-0}
   \\
   & =
   \sum_{m,l} \frac{\varrho_{m} - \varrho_{l}}{z + E_{m} - E_{l}} A_{ml}
   A_{lm}^{\dagger}.
   \label{eq:chi-ret-Lehmann-1}
\end{align}
In the last step, the integral converges for $z$ from the open upper half plane,
where $\chi^{\text{ret}}(z)$ is thus analytic. 

In this context we remark that the advanced exciton propagator 
\begin{align}
   \chi^{\text{adv}}(z) 
   & =
   \int_{-\infty}^{\infty}dt\, e^{izt} \chi^{\text{adv}}(t),
   \\
   \chi^{\text{adv}}(t) 
   & =
   i\Theta(-t)\left\langle \left[A(t)_{H},A^{\dagger}\right]\right\rangle 
\end{align}
satisfies $\chi^{\text{adv}}(z) = \chi^{\text{ret}}(z^{\ast})^{\ast}$ and is
analytic in the open lower half plane of $z$.  Its Lehmann representation is
formally identical to that of $\chi^{\text{ret}}(z)$ but is obtained from an
integral that converges only for $z$ from the open lower half plane.

The Lehmann representation of $\chi_{\beta}^{\text{Mat}}(\tau)$ in the case
$\tau>0$ is given by
\begin{equation}
   \chi_{\beta}^{\text{Mat}}(\tau)
   =
   -\sum_{m,l} \frac{e^{-\beta E_{m}}}{Z} e^{\tau(E_{m} - E_{l})} A_{ml}
   A_{lm}^{\dagger},
   \qquad \tau > 0.
\end{equation}
For $\chi_{\beta}^{\text{Mat}}(iX_{n})$ follows
\begin{equation}
   \chi_{\beta}^{\text{Mat}}(iX_{n})
   =
   \frac{1}{Z} \sum_{m,l} \frac{e^{-\beta E_{m}} - e^{-\beta E_{l}}}{iX_{n} + 
   E_{m} - E_{l}} A_{ml} A_{lm}^{\dagger},
\end{equation}
by straightforward integration and the use of $e^{i\beta X_{n}} = 1$.  This
coincides formally with the Lehmann representation of
$\chi_{\beta}^{\text{ret}}$ and $\chi_{\beta}^{\text{adv}}$ [compare
Eq.~(\ref{eq:chi-ret-Lehmann-1})].  However, $\chi_{\beta}^{\text{Mat}}(z)$ is
not defined on the upper or lower half plane of $z$ but only for the Matsubara
frequencies $z=iX_{n}$ on the imaginary $z$ axis. 

Next we consider $\chi_{\beta}^{\text{Mat}}$ in the limit of vanishing
temperature.  First we study the case $\epsilon_{\text{d}} <
\epsilon_{\text{d}0}$, in which $\ket{\Psi_{0}}$ is the ground state.  Due to
$e^{-\beta E_{m}}/Z \rightarrow \delta_{m,\Psi_{0}}$ for $\beta \rightarrow
\infty$, the Lehmann representation of $\chi_{\beta}^{\text{Mat}}(\tau)$
approaches 
\begin{equation}
   \chi_{\infty}^{\text{Mat}}(\tau)
   =
   -\Theta(\tau)\sum_{l} e^{\tau(E_{0} - E_{l})} A_{\Psi_{0}l} 
   A_{l\Psi_{0}}^{\dagger}.
\end{equation}
Here, $A_{l\Psi_{0}}^{\dagger}$ allows only for states $l$ with empty deep
level, such that $E_{l} \ge \bar{E}_{0} > E_{0}$. This means that all addends to
$\chi_{\infty}^{\text{Mat}}(\tau)$ decay as $e^{-\tau(\bar{E}_{0}-E_{0})}$ or
faster for $\tau\rightarrow\infty$.  Therefore, 
\begin{align}
   \chi_{\infty}^{\text{Mat}}(z)
   &=
   \int_{0}^{\infty}d\tau\, e^{z\tau} \chi_{\infty}^{\text{Mat}}(\tau)
   \label{eq:chi^Mat:tau-to-nu-1-0}
   \\
   &=
   \sum_{l} \frac{1}{z+E_{0}-E_{l}} A_{\Psi_{0}l} A_{l\Psi_{0}}^{\dagger}
   \label{eq:chi^Mat:tau-to-nu-1-1}
\end{align}
constitutes a convergent integral for all $z$ with $\Real z < \bar{E}_{0} -
E_{0} \, (>0)$. The very same expression results for
$\chi_{\beta}^{\text{ret}}(z) \rightarrow \chi_{\Psi_{0}}^{\text{ret}}(z) =
\chi_{\Psi_{0}}(z)$ from Eq.~(\ref{eq:chi-ret-Lehmann-1}) in the limit $\beta
\rightarrow \infty$, in which $\varrho_{m} \rightarrow \delta_{m\Psi_{0}}$.  The
two functions thus coincide on the intersection of their domains given by $\Real
z < \bar{E}_{0} - E_{0}$ and $\Imag z > 0$ (see
Fig.~\ref{fig:Domains-of-analyticity1}).

If instead $\epsilon_{\text{d}} > \epsilon_{\text{d}0}$ such that
$\nket{\bar{\Psi}_{0}}$ is the ground state, the Lehmann representation of
$\chi_{\beta}^{\text{Mat}}(\tau)$ approaches
\begin{equation}
   \chi_{\infty}^{\text{Mat}}(\tau)
   =
   -\Theta(-\tau) \sum_{m} e^{\tau(E_{m}-\bar{E}_{0})} A_{m\bar{\Psi}_{0}}
   A_{\bar{\Psi}_{0}m}^{\dagger}
\end{equation}
for $\beta \rightarrow \infty$. For $\tau \rightarrow -\infty$ the addends to
$\chi^{\text{Mat}}(\tau)$ decay as $e^{\tau(E_{0}-\bar{E}_{0})}$ or faster.
Therefore, 
\begin{align}
   \chi_{\infty}^{\text{Mat}}(z)
   &=
   \int_{-\infty}^{0}d\tau\, e^{z\tau} \chi_{\infty}^{\text{Mat}}(\tau)
   \label{eq:chi^Mat:tau-to-nu-2}
   \\
   &=
   \sum_{m} \frac{-1}{z+E_{m}-\bar{E}_{0}} A_{m\bar{\Psi}_{0}} 
   A_{\bar{\Psi}_{0}m}^{\dagger}
   \label{eq:chi^Mat:tau-to-nu-3}
\end{align}
constitutes a convergent integral for all $z$ with $\Real z > \bar{E}_{0} -
E_{0} \,(<0)$.  We remark in passing that the integration from $-\infty$ to $0$
in Eq.~(\ref{eq:chi^Mat:tau-to-nu-2}) results as the limit of the integration
from $-\beta$ to $0$.  In fact, it is possible to use this interval of
integration in Eq.~(\ref{eq:chi^Mat(iXn)}) instead of the one from $0$ to
$\beta$, due to $\chi_{\beta}^{\text{Mat}}(\tau-\beta) =
\chi_{\beta}^{\text{Mat}}(\tau)$ for $0 < \tau < \beta$ and due to
$e^{iX_{n}\beta} = 1$.

\begin{figure}
   \begin{center}
      \includegraphics[width=7cm]{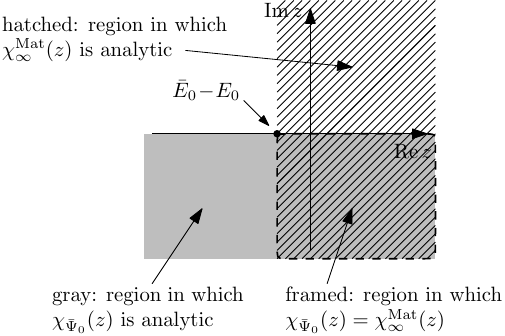}
      \caption{Domains of analyticity of $\chi_{\infty}^{\text{Mat}}(z)$ from
         Eq.~(\ref{eq:chi^Mat:tau-to-nu-2}) and $\chi_{\bar{\Psi}_{0}}(z) =
         \chi^{\text{adv}}_{\bar{\Psi}_{0}}(z)$ from
         Eq.~(\ref{eq:chi-adv-z-def}) in the case
         $\epsilon_{\text{d}}>\epsilon_{\text{d}0}$ in which
         $\nket{\bar{\Psi}_{0}}$ is the ground state.}
      \label{fig:Domains-of-analyticity2}
   \end{center}
\end{figure}
The very same expression as for $\chi_{\infty}^{\text{Mat}}(z)$ in
Eq.~(\ref{eq:chi^Mat:tau-to-nu-3}) results for $\chi_{\beta}^{\text{adv}}(z)$ in
the limit $\beta \rightarrow \infty$ with $\varrho_{m} \rightarrow
\delta_{m\bar{\Psi}_{0}}$.  In this limit, $\chi_{\beta}^{\text{adv}}(z)$
converges to
\begin{equation}
   \chi_{\bar{\Psi}_{0}}^{\text{adv}}(z)
   =
   \int_{-\infty}^{\infty}dt\, e^{izt} \chi_{\bar{\Psi}_{0}}^{\text{adv}}(t)
   \label{eq:chi-adv-z-def}
\end{equation}
with 
\begin{equation}
   \chi_{\bar{\Psi}_{0}}^{\text{adv}}(t)
   =
   i\Theta(-t)\bOk{\bar{\Psi}_{0}}{[A(t)_{H}, A^{\dagger}]}{\bar{\Psi}_{0}}.
\end{equation}
Due to $A^{\dagger}\nket{\bar{\Psi}_{0}} = 0$ the advanced exciton propagator
$\chi_{\bar{\Psi}_{0}}^{\text{adv}}$ is in turn identical to the time-ordered
propagator in the state $\ket{\bar{\Psi}_{0}}$:
\begin{equation}
   \chi_{\bar{\Psi}_{0}}^{\text{adv}}(t)
   =
   -i\bOk{\bar{\Psi}_{0}}{\TO A(t)_{H} A^{\dagger}}{\bar{\Psi}_{0}}
   =
   \chi_{\bar{\Psi}_{0}}(t).
\end{equation}
The two functions $\chi_{\infty}^{\text{Mat}}(z)$ and
$\chi_{\bar{\Psi}_{0}}(z)$ thus coincide on the intersection of their domains
given by $\Real z > \bar{E}_{0} - E_{0}$ and $\Imag z < 0$. This relation
between $\chi_{\infty}^{\text{Mat}}(z)$ and $\chi_{\bar{\Psi}_{0}}(z)$ is
sketched in Fig.~\ref{fig:Domains-of-analyticity2}.


\begin{widetext}

\section{Identity between Matsubara diagrams with limit propagators and diagrams
from zero-temperature formalism}
\label{sec:App-identification-Matsubara-0T}

In Secs.~\ref{subsec:Zero-temperature-limit} and~\ref{subsec:Setting-eps_d=0} we
affirmed that any Matsubara diagram for the particle-hole susceptibility that is
evaluated by using the zero-temperature limit propagator from
Eq.~(\ref{eq:Limit-prop-tau}) or (\ref{eq:Limit-prop-omega}) is equal to the
analytic continuation of the very same diagram evaluated in the zero-temperature
formalism.  This is shown in the following by using propagators in time
representation.  We expect that the result holds as well when the diagrams are
evaluated with propagators in frequency representation.

Let $D_{\text{lp}}^{\text{Mat}}(\tau)$ denote some time-dependent diagram with
$n$ interaction vertices that contributes to the particle-hole susceptibility.
As the index ``lp'' indicates, the diagram is evaluated with limit propagators
[in contrast to diagrams $D_\infty^{\text{Mat}}(\tau) =
\lim_{\beta\rightarrow\infty} D_{\beta}^{\text{Mat}}(\tau)$ which are evaluated
with finite-temperature propagators, the limit $\beta\rightarrow\infty$ being
taken afterwards].  To each vertex in $D_{\text{lp}}^{\text{Mat}}(\tau)$ is
attached an incoming and an outgoing deep-state line as well as an incoming and
an outgoing conduction-state line, the lines representing Hartree dressed limit
propagators.  In total there are $(n+1)$ deep-state lines and $(n+1)$
conduction-state lines.  Since the deep-state lines
$G_{\infty,\text{d}}^{\text{Mat}}(\tau) = e^{-\tilde{\epsilon}_{\text{d}}\tau}
\Theta(-\tau)$ are directed backwards in imaginary time they are concatenated in
a single sequence from $\tau$ to $0$, connecting the vertices in a unique time
order.  Accordingly, we choose the time labels $\tau_{j}$ of the vertices such
that $\tau \ge \tau_{n} \ge \dots \ge \tau_{1} \ge 0$.  Then the product of all
deep-state propagators is
\begin{equation}
   e^{-\tilde{\epsilon}_{\text{d}}(\tau_{n}-\tau)}
   e^{-\tilde{\epsilon}_{\text{d}}(\tau_{n-1}-\tau_{n})}
   \dots
   e^{-\tilde{\epsilon}_{\text{d}}(\tau_{1}-\tau_{2})}
   e^{-\tilde{\epsilon}_{\text{d}}(0-\tau_{1})}
   =
   e^{\tilde{\epsilon}_{\text{d}}\tau}.
\end{equation}
 
\begin{figure}
   \begin{center}
      \includegraphics[width=3.5cm]{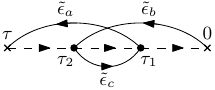}
   \end{center}
   \caption{Example for a diagram $D_{\text{lp}}^{\text{Mat}}(\tau)$ of order
      $n=2$.  The (Hartree renormalized) energies of the conduction-state lines
      are $\tilde{\epsilon}_{a} > 0$, $\tilde{\epsilon}_{b} > 0$ and
      $\tilde{\epsilon}_{c} < 0$ and are addressed as
      $\tilde{\epsilon}^{\text{in}} = \tilde{\epsilon}_{1}^{\text{out}} =
      \tilde{\epsilon}_{a}$, $\tilde{\epsilon}_{1}^{\text{in}} =
      \tilde{\epsilon}_{2}^{\text{out}} = \tilde{\epsilon}_{c}$,
      $\tilde{\epsilon}_{2}^{\text{in}} = \tilde{\epsilon}_{b}$.}
   \label{fig:example-for-eps-in-out}
\end{figure}

The vertices are connected in some way by the conduction-state lines. Due to the
unique time order of the vertices, the conduction-state lines can be grouped
into lines of hole propagation with $\tilde{\epsilon}_{k} < 0$ and lines of
particle propagation with $\tilde{\epsilon}_{k} > 0$.  Let $n_{\text{h}}$ be the
number of conduction-state hole propagators. The product of all conduction-state
lines is then given by
\begin{equation}
   (-1)^{n+1} (-1)^{n_{\text{h}}} e^{-\tau\tilde{\epsilon}^{\text{in}}}
   e^{-\tau_{n}(\tilde{\epsilon}_{n}^{\text{in}} - 
   \tilde{\epsilon}_{n}^{\text{out}})} \dots
   e^{-\tau_{1}(\tilde{\epsilon}_{1}^{\text{in}} - 
   \tilde{\epsilon}_{1}^{\text{out}})},
\end{equation}
where $\tilde{\epsilon}_{j}^{\text{in}}$ and $\tilde{\epsilon}_{j}^{\text{out}}$
denote the (Hartree renormalized) energy $\tilde{\epsilon}$ of that
conduction-state propagator which enters or leaves, respectively, the vertex
with time $\tau_{j}$.  Furthermore, $\tilde{\epsilon}^{\text{in}}$ denotes the
energy of the conduction-state propagator ending at time $\tau$.  An example for
this labeling is given in Fig.~\ref{fig:example-for-eps-in-out}.

The frequency dependence of the diagram is thus given by
\begin{align}
   D_{\text{lp}}^{\text{Mat}}(z) 
   & =
   \frac{1}{V} \int_{0}^{\infty}d\tau\, e^{z\tau} 
   (-1)^{n_{\text{loop}}}
   \left(-\frac{U}{V}\right)^{n}
   \int_{0}^{\tau} d\tau_{n} \int_{0}^{\tau_{n}} d\tau_{n-1} \dots 
   \int_{0}^{\tau_{2}}d\tau_{1}
   \nonumber \\
   & \quad\;
   \times \rho^{n+1} \int_{-\xi_{0}}^{0} d\tilde{\epsilon}_{1} \dots 
   \int_{-\xi_{0}}^{0}d\tilde{\epsilon}_{n_{\text{h}}} 
   \int_{0}^{\xi_{0}}d\tilde{\epsilon}_{n_{\text{h}}+1} \dots
   \int_{0}^{\xi_{0}}d\tilde{\epsilon}_{n+1}
  \nonumber \\
   & \quad\;
   \times e^{\tilde{\epsilon}_{\text{d}}\tau} (-1)^{n+1} (-1)^{n_{\text{h}}}
   e^{-\tau\tilde{\epsilon}^{\text{in}}} 
   e^{-\tau_{n}(\tilde{\epsilon}_{n}^{\text{in}} 
   - \tilde{\epsilon}_{n}^{\text{out}})} \dots
   e^{-\tau_{1}(\tilde{\epsilon}_{1}^{\text{in}}
   -\tilde{\epsilon}_{1}^{\text{out}})}
   \\
   & =
   (-1)^{n_{\text{h}}+1} (-1)^{n_{\text{loop}}}\frac{U^{n}\rho^{n+1}}{V^{n+1}}
   \int_{-\xi_{0}}^{0} d\tilde{\epsilon}_{1} \dots
   \int_{-\xi_{0}}^{0} d\tilde{\epsilon}_{n_{\text{h}}}
   \int_{0}^{\xi_{0}}d\tilde{\epsilon}_{n_{\text{h}}+1} \dots
   \int_{0}^{\xi_{0}}d\tilde{\epsilon}_{n+1}
   \nonumber \\
   & \quad\;
   \times \int_{0}^{\infty}d\tau\, 
   e^{\tau(z+\tilde{\epsilon}_{\text{d}}-\tilde{\epsilon}^{\text{in}})}
   \int_{0}^{\tau}d\tau_{n}\, e^{-\tau_{n}(\tilde{\epsilon}_{n}^{\text{in}} 
   - \tilde{\epsilon}_{n}^{\text{out}})} \dots
   \int_{0}^{\tau_{2}}d\tau_{1}\, e^{-\tau_{1}(\tilde{\epsilon}_{1}^{\text{in}}
   - \tilde{\epsilon}_{1}^{\text{out}})}
   \label{eq:D^Mat_lp-step1}
\end{align}
[compare Eqs.~(\ref{eq:chi-from-GF}) and~(\ref{eq:diagram-rule})]; due to the
special role of the deep-state line, the prefactors $(-1)^P/(2^{n_{\text{eq}}}
S)$ from Eq.~(\ref{eq:diagram-rule}) do not arise here. In
Eq.~(\ref{eq:D^Mat_lp-step1}), the rightmost integral over imaginary time yields
\begin{equation}
   \int_{0}^{\tau_{2}}d\tau_{1}\, e^{-\tau_{1}(\tilde{\epsilon}_{1}^{\text{in}}
   - \tilde{\epsilon}_{1}^{\text{out}})}
   =
   - \frac{e^{-\tau_{2}(\tilde{\epsilon}_{1}^{\text{in}} 
   - \tilde{\epsilon}_{1}^{\text{out}})} - 1}{\tilde{\epsilon}_{1}^{\text{in}}
   - \tilde{\epsilon}_{1}^{\text{out}}}.
   \label{eq:D^Mat_lp-first-integral}
\end{equation}
This together with the next integral yields
\begin{equation}
   (-1)^{2} \left[\frac{e^{-\tau_{3}(\tilde{\epsilon}_{2}^{\text{in}} 
   - \tilde{\epsilon}_{2}^{\text{out}} + \tilde{\epsilon}_{1}^{\text{in}}
   - \tilde{\epsilon}_{1}^{\text{out}})} - 1}{\tilde{\epsilon}_{2}^{\text{in}}
   - \tilde{\epsilon}_{2}^{\text{out}} + \tilde{\epsilon}_{1}^{\text{in}} 
   - \tilde{\epsilon}_{1}^{\text{out}}} 
   - \frac{e^{-\tau_{3}(\tilde{\epsilon}_{2}^{\text{in}} 
   - \tilde{\epsilon}_{2}^{\text{out}})} - 1}{\tilde{\epsilon}_{2}^{\text{in}}
   - \tilde{\epsilon}_{2}^{\text{out}}}\right] 
   \frac{1}{\tilde{\epsilon}_{1}^{\text{in}} 
   - \tilde{\epsilon}_{1}^{\text{out}}},
   \label{eq:next-time-integral}
\end{equation}
and so on. In particular, the first $n$ integrals produce each a prefactor
$(-1)$.  Finally, the last imaginary-time integral consists of addends that are
made of products of energy denominators and integrations of the type
\begin{equation}
   \int_{0}^{\infty} d\tau\, e^{\tau(z + \tilde{\epsilon}_{\text{d}} + E)}
   =
   \left.\frac{e^{\tau(z + \tilde{\epsilon}_{\text{d}} + E)}}{z 
   +\tilde{\epsilon}_{\text{d}} + E}\right|_{0}^{\infty}
   =
   -\frac{1}{z + \tilde{\epsilon}_{\text{d}} + E},
   \label{eq:D^Mat_lp-last-integral}
\end{equation}
where $E$ is one of
\begin{gather}
   -\tilde{\epsilon}^{\text{in}},
   \\
   -\tilde{\epsilon}^{\text{in}} + (\tilde{\epsilon}_{n}^{\text{out}}
   - \tilde{\epsilon}_{n}^{\text{in}}),\\
   -\tilde{\epsilon}^{\text{in}} + (\tilde{\epsilon}_{n}^{\text{out}}
   - \tilde{\epsilon}_{n}^{\text{in}}) + (\tilde{\epsilon}_{n-1}^{\text{out}}
   - \tilde{\epsilon}_{n-1}^{\text{in}}),\\
   \vdots \nonumber\\
   -\tilde{\epsilon}^{\text{in}} + (\tilde{\epsilon}_{n}^{\text{out}}
   - \tilde{\epsilon}_{n}^{\text{in}}) + (\tilde{\epsilon}_{n-1}^{\text{out}}
   - \tilde{\epsilon}_{n-1}^{\text{in}}) + \dots 
   + (\tilde{\epsilon}_{1}^{\text{out}} - \tilde{\epsilon}_{1}^{\text{in}}).
\end{gather}
The integral in Eq.~(\ref{eq:D^Mat_lp-last-integral}) converges for $\Real z <
-\tilde{\epsilon}_{\text{d}}$, as discussed below. This last integral produces
again a prefactor $(-1)$. The combined prefactor $(-1)^{n+1}$ from the time
integrals can be merged with the other prefactors to yield
\begin{multline}
   D_{\text{lp}}^{\text{Mat}}(z)
   =
   (-1)^{n+n_{\text{h}}} (-1)^{n_{\text{loop}}} \frac{U^{n}\rho^{n+1}}{V^{n+1}}
   \int_{-\xi_{0}}^{0} d\tilde{\epsilon}_{1} \dots 
   \int_{-\xi_{0}}^{0} d\tilde{\epsilon}_{n_{\text{h}}}
   \int_{0}^{\xi_{0}} d\tilde{\epsilon}_{n_{\text{h}}+1} \dots
   \int_{0}^{\xi_{0}} d\tilde{\epsilon}_{n+1} 
   \sum_j \frac{1}{z + \tilde{\epsilon}_{\text{d}} + E_j} \dots
   \frac{1}{\tilde{\epsilon}_{1}^{\text{in}} 
   - \tilde{\epsilon}_{1}^{\text{out}}}
   \label{eq:D^Mat_lp-value}
\end{multline}
as analytic function on the region given by $\Real z <
-\tilde{\epsilon}_{\text{d}}$.  The sum on the very right of this expression
runs over all resulting combinations $j$ of factors.  The $(n-1)$ inner factors
represented by the dots close to the end of Eq.~(\ref{eq:D^Mat_lp-value}) may
carry additional sign factors depending on their precise provenance from the
inner time integrations [compare Eq.~(\ref{eq:next-time-integral})].

\begin{figure}
   \begin{center}
      \includegraphics{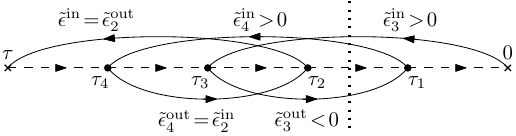}
   \end{center}
   \caption{Example for a diagram $D_{\text{lp}}^{\text{Mat}}(\tau)$ of order
      $n=4$ and for the case $m=2$.  The dotted vertical line separates the
      block of (internal and external) vertices with times $\tau, \tau_{n},
      \dots, \tau_{n-m}$ from the block with times $\tau_{n-m-1}, \dots,
      \tau_{1},0$.}
      \label{fig:example-for-blocks}
\end{figure}

In order to scrutinize the convergence of the outermost imaginary-time
integration from Eq.~(\ref{eq:D^Mat_lp-last-integral}), consider the general
case
\begin{equation}
   E
   =
   -\tilde{\epsilon}^{\text{in}} + (\tilde{\epsilon}_{n}^{\text{out}}
   - \tilde{\epsilon}_{n}^{\text{in}}) + (\tilde{\epsilon}_{n-1}^{\text{out}}
   - \tilde{\epsilon}_{n-1}^{\text{in}}) + \dots
   + (\tilde{\epsilon}_{n-m}^{\text{out}} - \tilde{\epsilon}_{n-m}^{\text{in}}).
\end{equation}
On the right-hand side of this equation, the energies of all conduction-state
lines connecting the vertices at times $\tau, \tau_{n}, \dots, \tau_{n-m}$ among
each other cancel out, as they appear once as some
$\tilde{\epsilon}_{i}^{\text{in}}$ and once as some
$\tilde{\epsilon}_{j}^{\text{out}}$ (see Fig.~\ref{fig:example-for-blocks} for
an example).  There remains the sum of the energies of conduction-state lines
connecting the block of vertices at times $\tau, \tau_{n}, \dots, \tau_{n-m}$
with the block of vertices at times $\tau_{n-m-1}, \dots, \tau_{1},0$. If such a
line is directed from the former block to the latter one, then it is a hole line
with negative energy which enters $E$ as some
$\tilde{\epsilon}_{j}^{\text{out}}$.  If the line runs in the opposite
direction, then it is a particle line with positive energy which enters $E$ as
some $-\tilde{\epsilon}_{j}^{\text{in}}$. Consequently, $E \le 0$.  This means
that the convergence of the integral in Eq.~(\ref{eq:D^Mat_lp-last-integral}) is
guaranteed if $\Real z < -\tilde{\epsilon}_{\text{d}}$.

We note that the resulting domain of analyticity differs from the domain of
analyticity of $\chi_{\infty}^{\text{Mat}}(z)$ which is given by $\Real z <
\bar{E}_{0} - E_{0}\, (=\epsilon_{\text{d}0} - \epsilon_{\text{d}} =
\tilde{\epsilon}_{\text{d}0} - \tilde{\epsilon}_{\text{d}})$ for
$\epsilon_{\text{d}} < \epsilon_{\text{d}0}$ and by $\Real z > \bar{E}_{0} -
E_{0}$ for $\epsilon_{\text{d}} > \epsilon_{\text{d}0}$ (see
Sec.~\ref{subsec:Linear-response-rate} and
Appendix~\ref{sec:App-analytic-behaviour}).  
This discrepancy highlights that there are diagrams (those with self-energy type
insertions on the lines) for which the transition to limit propagators is not
correct.

Now we switch from imaginary times to real times and consider the diagram as a
contribution to $\chi_{\Psi_{0}}(z)$ in the state $\ket{\Psi_{0}}$ as determined
by the real-time zero-temperature formalism.  We will find the same diagram
value on the overlap of the corresponding domains.  The Hartree dressed
zero-temperature propagators are given by
\begin{equation}
   G(t) 
   =
   ie^{-i\tilde{\epsilon}t} \left[\Theta(-\tilde{\epsilon})
   -\Theta(t)\right].
\end{equation}
Here, the exponent $-i \tilde \epsilon t$ is understood as limit of $-i
(1-i\eta) \tilde \epsilon t$ for $\eta \rightarrow 0^+$ (compare
Ref.~\cite{Diekmann2021}).  Again, the deep-state propagator is directed
backwards in time, establishing a unique time order of the vertices with $t \ge
t_{n} \ge \dots \ge t_{1} \ge 0$.  Similar to the Matsubara case above, the
product of all deep-state lines is $i^{n+1} e^{i\tilde{\epsilon}_{\text{d}}t}$,
and the product of all conduction-state lines is
\begin{equation}
   (-i)^{n+1} (-1)^{n_{\text{h}}} 
   e^{-it\tilde{\epsilon}^{\text{in}}}
   e^{-it_{n}(\tilde{\epsilon}_{n}^{\text{in}}
   -\tilde{\epsilon}_{n}^{\text{out}})} \dots
   e^{-it_{1}(\tilde{\epsilon}_{1}^{\text{in}}
   -\tilde{\epsilon}_{1}^{\text{out}})}.
\end{equation}
The contribution of the diagram to $\chi_{\Psi_{0}}(z)$ thus turns out to be
\begin{multline}
   D_{\Psi_{0}}(z) 
   =
   i^{n+1} (-1)^{n_{\text{h}}+1} (-1)^{n_{\text{loop}}} 
   \frac{U^{n} \rho^{n+1}}{V^{n+1}}
   \int_{-\xi_{0}}^{0} d\tilde{\epsilon}_{1} \dots
   \int_{-\xi_{0}}^{0} d\tilde{\epsilon}_{n_{\text{h}}}
   \int_{0}^{\xi_{0}}d\tilde{\epsilon}_{n_{\text{h}}+1} \dots
   \int_{0}^{\xi_{0}}d\tilde{\epsilon}_{n+1}
   \\
   \times\int_{0}^{\infty} dt\, e^{it(z+\tilde{\epsilon}_{\text{d}}
   -\tilde{\epsilon}^{\text{in}})}
   \int_{0}^{t}dt_{n}\, e^{-it_{n}(\tilde \epsilon_{n}^{\text{in}}
   -\tilde{\epsilon}_{n}^{\text{out}})} \dots
   \int_{0}^{t_{2}}dt_{1}\, e^{-it_{1}(\tilde \epsilon_{1}^{\text{in}}
   -\tilde{\epsilon}_{1}^{\text{out}})}.
\end{multline}
The real-time integrations at the end of this expression yield precisely the
same results as the corresponding imaginary-time integrations in
Eqs.~(\ref{eq:D^Mat_lp-first-integral}--\ref{eq:D^Mat_lp-last-integral}) above,
with the single exception that the prefactor $(-1)$ resulting from every
integral is replaced by a factor $i$.  The combined factor $i^{n+1}$ can be
merged with the other prefactors to yield for $D_{\Psi_{0}}(z)$ the identical
expression as indicated for $D_{\text{lp}}^{\text{Mat}}(z)$ in
Eq.~(\ref{eq:D^Mat_lp-value}).  In the case of $D_{\Psi_{0}}(z)$, the final time
integration decays into addends of the type
\begin{equation}
   \int_{0}^{\infty}dt\, e^{it(z+\tilde{\epsilon}_{\text{d}}+E)},
\end{equation}
which converge for $z$ from the upper half plane, as opposed to the condition
$\Real z < -\tilde{\epsilon}_{\text{d}}$ for $D_{\text{lp}}^{\text{Mat}}(z)$. On
the quadrant allowed by both conditions, the values are identical. This allows
for an analytic continuation from the limit-propagator Matsubara diagram to the
one from the zero-temperature formalism.

\end{widetext}


\section{Illustration of the identity between $\chi_{\infty}^{\text{Mat}}(iX)$
and $\chi_{\Psi_{0}}(X)$}
\label{sec:App-illustrate-identity}

\begin{figure*}
   \begin{center}
      \includegraphics{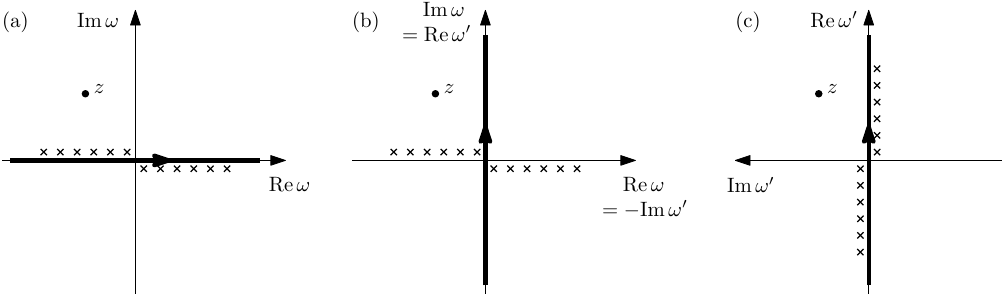}
   \end{center}
   \caption{(a) Integration contour and poles of the integrand in
      Eq.~(\ref{eq:chi-non-int-poles-1}).  The crosses symbolize the numerous
      poles at $\tilde{\epsilon}_{k}-i\eta\sgn\tilde{\epsilon}_{k}$ stemming
      from the conduction band. (b) Rotated integration contour and the poles of
      the integrand in Eq.~(\ref{eq:chi-non-int-poles-2}). (c) Integration
      contour and the poles of the integrand in
      Eq.~(\ref{eq:chi-non-int-poles-3}).}
   \label{fig:chi-non-int-poles}
\end{figure*}

In order to illustrate the remarkable equality between the leading-logarithmic
approximations for $\chi_\infty^{\text{Mat}}(iX)$ and $\chi_{\Psi_{0}}(X)$ that
was described in Sec.~\ref{sec:identity-leading-log-real-imag}, we refer to the
simple example of the noninteracting particle-hole susceptibility and the
corresponding distribution of poles in the complex plane.  The noninteracting
particle-hole susceptibility is given in the real-time zero-temperature
formalism by
\begin{multline}
   \chi_{\Psi_{0}}^{\text{nonint}}(z)
   =
   -i \frac{\rho}{V} \int_{-\xi_{0}}^{\xi_{0}} d\tilde{\epsilon}
   \int_{-\infty}^{\infty} \frac{d\omega}{2\pi}
   \frac{1}{\omega-z-i\eta} 
   \\ \times
   \frac{1}{\omega-\tilde{\epsilon}+i\eta\sgn\tilde{\epsilon}}.
   \label{eq:chi-non-int-poles-1}
\end{multline}

The contour of $\omega$ integration and the poles of the integrand in the
complex $\omega$ plane are shown in Fig.~\ref{fig:chi-non-int-poles}(a), with
$z$ chosen in the open upper left quadrant.  The integration contour can now be
rotated:  The integral along the negative real $\omega$ axis is identical to the
integral along the negative imaginary $\omega$ axis as there are no
nonanalyticities in the lower left quadrant and as the integrand decays as
$1/\omega^{2}$ for $|\omega|\rightarrow\infty$.  Due to the same reasons, the
integral along the positive real $\omega$ axis equals the integral along the
positive imaginary $\omega$ axis.  With $\omega=i\omega'$ we thus obtain
\begin{align}
   \chi_{\Psi_{0}}^{\text{nonint}}(z)
   &=
   \frac{\rho}{V} \int_{-\xi_{0}}^{\xi_{0}} d\tilde{\epsilon}
   \int_{-\infty}^{\infty} \frac{d\omega'}{2\pi} \frac{1}{i\omega'-z}
   \frac{1}{i\omega'-\tilde{\epsilon}}
   \label{eq:chi-non-int-poles-2}
   \\
   &=
   \chi_\infty^{\text{Mat, nonint}}(z).
\end{align}
The infinitesimal shifts $\pm i\eta$ can be omitted once the integration is
along the imaginary axis.  The new integration contour is shown in
Fig.~\ref{fig:chi-non-int-poles}(b).  

So far we rederived the analytic continuation between $\chi_{\Psi_{0}}$ and
$\chi_\infty^{\text{Mat}}$ for the noninteracting case.  But now we rewrite
$\chi_\infty^{\text{Mat, nonint}}(z)$ as
\begin{equation}
   \chi_\infty^{\text{Mat, nonint}}(z)
   =
   -i \int_{-\infty}^{\infty} \frac{d\omega'}{2\pi} \frac{1}{\omega'-(-iz)}
   \frac{\rho}{V} \int_{-\xi_{0}}^{\xi_{0}} 
   \frac{d\tilde{\epsilon}}{i\omega'-\tilde{\epsilon}}
\end{equation}
and use 
\begin{align}
   \frac{\rho}{V} \int_{-\xi_{0}}^{\xi_{0}} 
   \frac{d\tilde{\epsilon}}{i\omega'-\tilde{\epsilon}}
   &\approx
   -i \pi \frac{\rho}{V} \sgn(\omega') \Theta(\xi_{0}-|\omega'|)
   \label{eq:local-cond-el-prop-double-approx}
   \\
   &\approx
   \frac{\rho}{V} \int_{-\xi_{0}}^{\xi_{0}} 
   \frac{d\tilde{\epsilon}}{\omega'-\tilde{\epsilon} 
   + i\eta\sgn\tilde{\epsilon}}.
\end{align}
Here, the first approximation is precisely
Eq.~(\ref{eq:local-cond-el-prop-approx}); the second one is known from
Ref.~\cite{Roulet1969} to respect the leading-logarithmic
contributions to $\chi_{\Psi_{0}}$.  We obtain
\begin{align}
   &\chi_\infty^{\text{Mat, nonint}}(z)
   \nonumber
   \\
   &\approx
   -i\frac{\rho}{V} \int_{-\xi_{0}}^{\xi_{0}} d\tilde{\epsilon}
   \int_{-\infty}^{\infty} \frac{d\omega'}{2\pi} \frac{1}{\omega'-(-iz)-i\eta}
   \frac{1}{\omega'-\tilde{\epsilon} + i\eta\sgn\tilde{\epsilon}}
   \label{eq:chi-non-int-poles-3}
   \\
   &= \chi_{\Psi_{0}}^{\text{nonint}}(-iz)
   \end{align}
and in particular $\chi_\infty^{\text{Mat, nonint}}(iX) \approx
\chi_{\Psi_{0}}^{\text{nonint}}(X)$. We note that for $z$ from the open upper
left quadrant we could simply reintroduce the infinitesimal shift $(-i\eta)$.
The poles of the complex integrand are shown in
Fig.~\ref{fig:chi-non-int-poles}(c). In comparison to
Fig.~\ref{fig:chi-non-int-poles}(a) it becomes apparent that the main effect of
Eq.~(\ref{eq:local-cond-el-prop-double-approx}), i.e., of
approximation~(\ref{eq:local-cond-el-prop-approx}), is to reestablish the pole
structure of the conduction-electron states as known from the real-time
zero-temperature formalism in a coordinate system with interchanged real- and
imaginary-frequency axes.


\bibliography{leading-log-Matsubara-FRG}

\end{document}